\pdfoutput=1
\documentclass[cits]{PoS}

\def\etal{{\it et al.}}
\newcommand{\ud}{\ensuremath{\mathrm{d}}}
\newcommand{\dms}{\ensuremath{\Delta m_{s}}}
\newcommand{\DGs}{\ensuremath{\Delta\Gamma_{s}}}
\newcommand{\Gs}{\ensuremath{\Gamma_{s}}}

\newcommand{\delperp}{\ensuremath{\delta_{\perp}}}
\newcommand{\delpar}{\ensuremath{\delta_{\|}}}
\newcommand{\delzero}{\ensuremath{\delta_0}}

\newcommand{\rmS}{\ensuremath{\mathrm{S}}}

\newcommand{\Bsb}{\ensuremath{\overline{B}_s^0}}

\usepackage{graphicx}
\usepackage{braket}
\usepackage{amsmath}
\usepackage{afterpage}

\title{New physics from flavour }

\ShortTitle{New physics from flavour}

\author{\speaker{Sheldon Stone}\thanks{Work supported by U.S. National Science Foundation}\\
        Physics Dept., 201 Physics Building, Syracuse University, Syracuse, NY, USA, 13244-1130\\
        E-mail: \email{stone@physics.syr.edu}}

\abstract{I report on investigations of physics beyond the Standard Model using experiments designed to investigate $b$ quark and $c$ quark interactions. Generalized searches involving loop diagrams using heavy meson mixing and $CP$ violation experiments imply limits on New Physics (NP) at large scales from 100-100,000 TeV. There are  ways of avoiding these limits, but satisfying their requirements puts severe restrictions on NP properties. Specific constraints on NP models  are discussed from individual measurements including $b\to s\gamma$, $\Bsb\to\mu^+\mu^-$,  $\overline{B}^0\to \overline{K}^*\ell^+\ell^-$, and  $CP$ violation in the $B^0$ and $B_s^0$ systems. Hints of deviations from  Standard Model predictions from other decays including $B^-\to \tau^-\overline{\nu}$, and $\overline{B}\to D^{(*) } \tau^-\overline{\nu}$ are investigated. Finally, searches for dark matter and Majorana neutrinos are reviewed.}

\FullConference{36th International Conference on High Energy Physics,\\
		July 4-11, 2012\\
		Melbourne, Australia}

\begin{document}

\section{Introduction: Reasons for physics beyond the Standard Model}

Although the Standard Model (SM) of particle physics provides an excellent description of electroweak and strong interactions,  there are many reasons that we expect to observe new forces giving rise to new particles at larger masses than the known fermions or bosons. One oft noted source of this belief is the observation of dark matter in the cosmos as evidenced by galactic angular velocity distributions \cite{Zwicky}, gravitational lensing \cite{lensing}, and galactic collisions \cite{galactic}. The existence of dark energy, believed to cause the accelerating expansion of the Universe, is another source of mystery \cite{darkenergy}. The fine tuning of quantum corrections needed to keep, for example, the Higgs boson mass at the electroweak scale rather than near the Planck scale is another reason habitually mentioned for new physics (NP) and is usually called ``the hierarchy problem'' \cite{heir}.

\begin{figure} [hbt]
\begin{center}
\includegraphics[width=.6\textwidth]{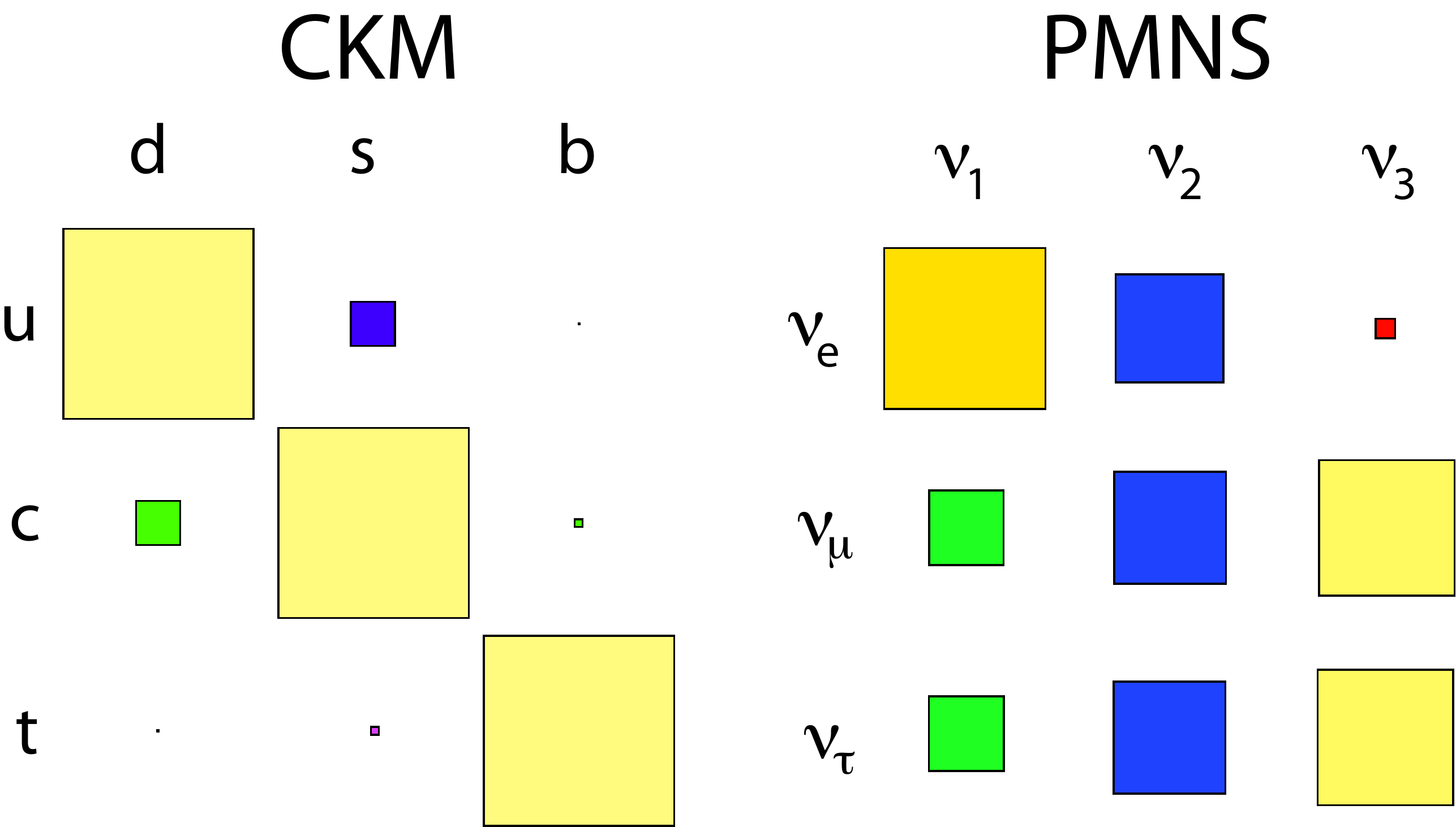} 
\caption{(left) Sizes of the the CKM matrix elements for quark mixing, and (right) the PMNS matrix elements for neutrino mixing. The area of the squares represents the square of the matrix elements.} \label{CKM-PMNS} 
\end{center}
\end{figure}

It is interesting to note that the above cited reasons are all tied in one way or another to gravity. Dark matter may or may not have purely gravitational interactions, dark energy may be explained by a cosmological constant or at least be a purely general relativistic phenomena, and the Planck scale is defined by gravity; other scales may exist at much lower energies, so the quantum corrections could be much smaller. There are, however, many observations that are not explained by the SM, and have nothing to do with gravity, as far as we know. Consider the size of the quark mixing matrix (CKM) elements \cite{Nakao} and also the neutrino mixing matrix (PMNS) elements \cite{PDG}. These are shown pictorially in Fig.~\ref{CKM-PMNS}.  We do not understand the relative sizes of these values or nor the relationship between quarks and neutrinos.

\begin{figure} [hbt]
\begin{center}
\includegraphics[width=.6\textwidth]{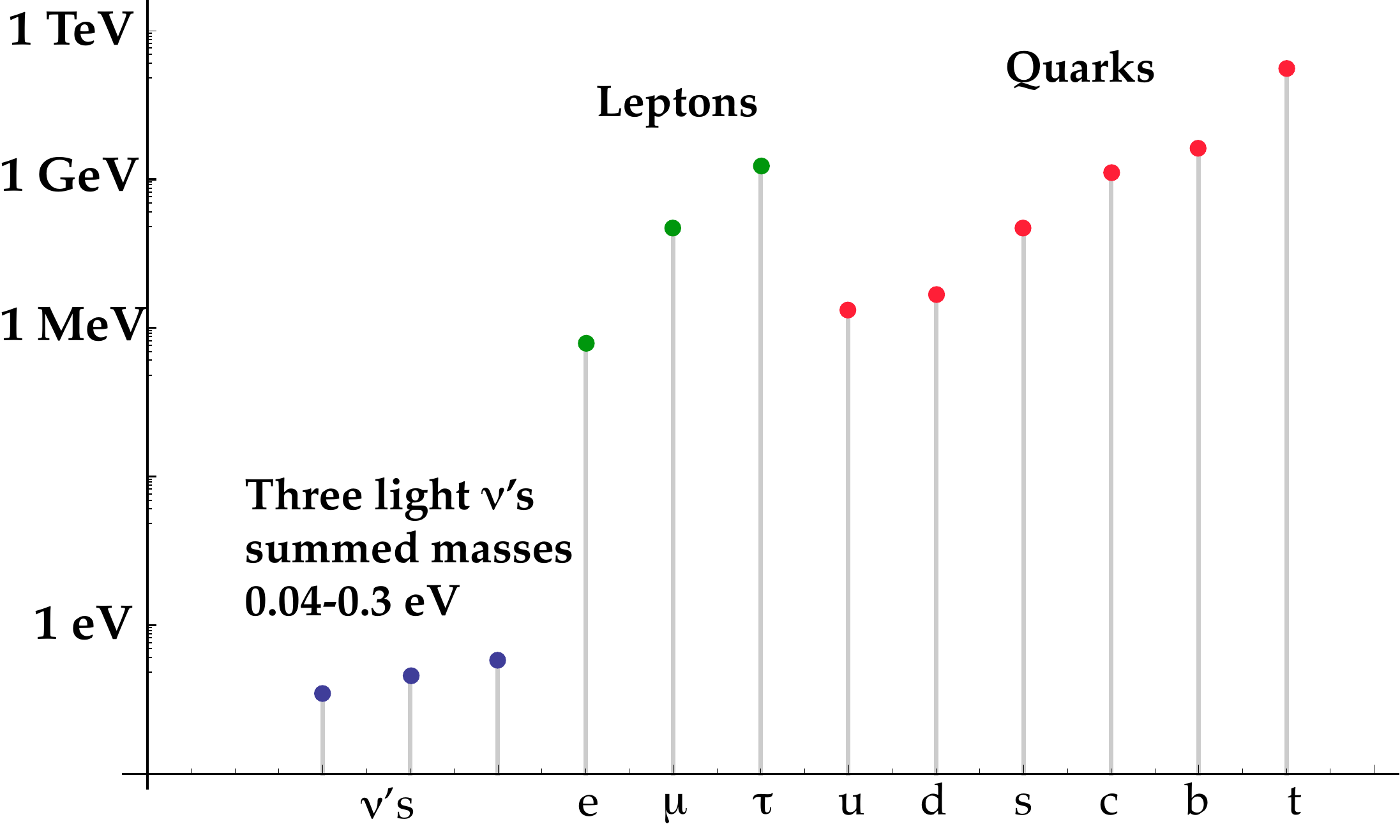} 
\caption{Lepton and quark masses.} \label{Masses} 
\end{center}
\end{figure}

We also do not understand the masses of the fundamental matter constituents, the quarks and leptons. Not only are they not predicted, but also the relationships among them are not understood. These masses, shown in Fig.~\ref{Masses}, span 12 orders of magnitude \cite{PDG}. There may be a connections between the mass values and the values of the mixing matrix elements, but thus far no connection besides simple numerology exists.

What we are seeking is a new theoretical explanation of the above mentioned facts. Of course, any new model must explain all the data, so that any one measurement could confound a model. It is not a good plan, however, to try and find only one discrepancy; experiment must determine a consistent pattern of deviations to restrict possible theoretical explanations, and to be sure the new phenomena is real. 

\section{Use of flavour physics as a new physics discovery tool}
\label{sec:NP}
While measurements of CKM parameters and masses are important, as they are fundamental constants of nature, the main purpose of flavour physics is to find and/or define the properties of physics beyond the SM.
Flavour physics probes large mass scales via virtual quantum loops. Examples of the importance of such loops are changes in the $W$ boson mass ($M_W$) from the existence of the $t$ quark of mass $m_t$, $dM_W\propto m_t^2$, and changes due to the  existence of the Higgs boson of mass $M_H$, $dM_W\propto {\rm ln}(M_H)$.

\begin{figure} [h!]
\begin{center}
\includegraphics[width=.8\textwidth]{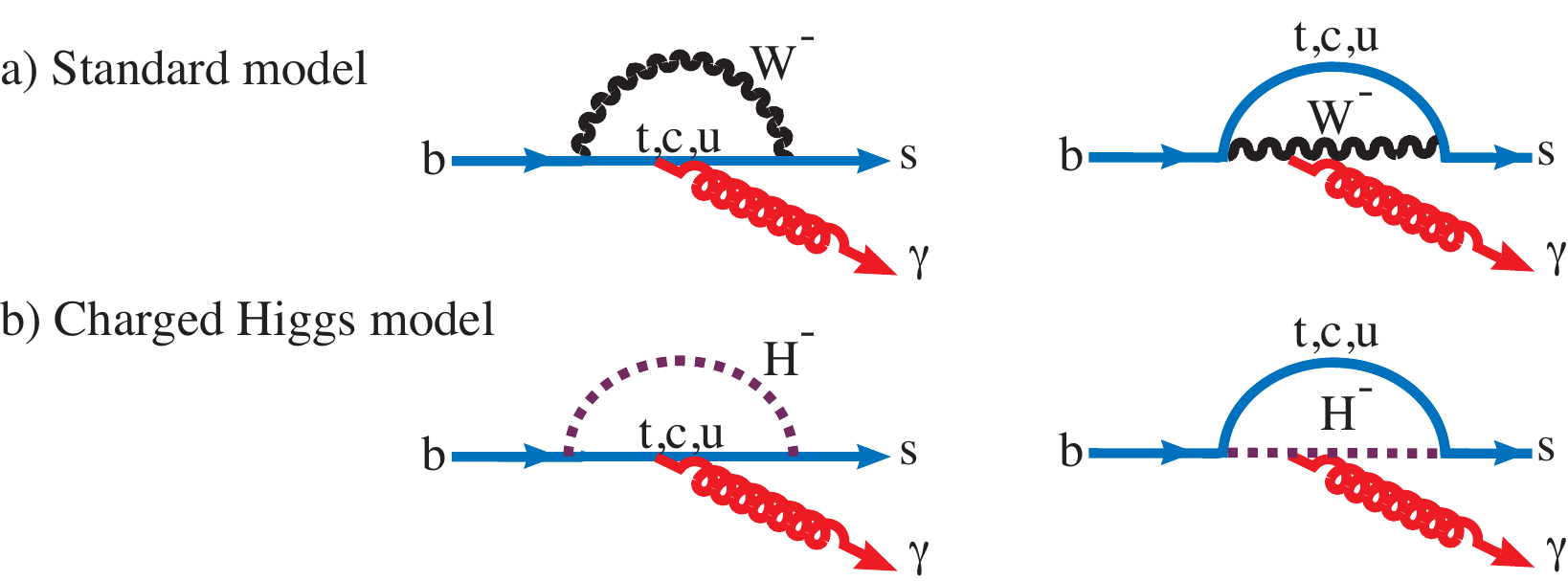} 
\caption{(a) SM diagrams for the quark level process $b\to s\gamma$ and (b) NP diagrams mediated by a charged Higgs boson.} \label{ChargedHiggs-penguin} 
\end{center}
\end{figure}

Strong constraints on NP are provided by individual processes. Each process provides a different constraint. Consider for example the inclusive decay $b\to s\gamma$. The SM diagrams are shown in Fig.~\ref{ChargedHiggs-penguin}(a). The SM calculation considers either a $B^-$ or $\overline{B}^0$ meson and then sums over all decays where a hard photon emerges. The Feynman diagrams corresponding to a NP process with a virtual charged Higgs boson are shown in Fig.~\ref{ChargedHiggs-penguin}(b).

\begin{figure} [hbt]
\begin{center}
\includegraphics[width=.6\textwidth]{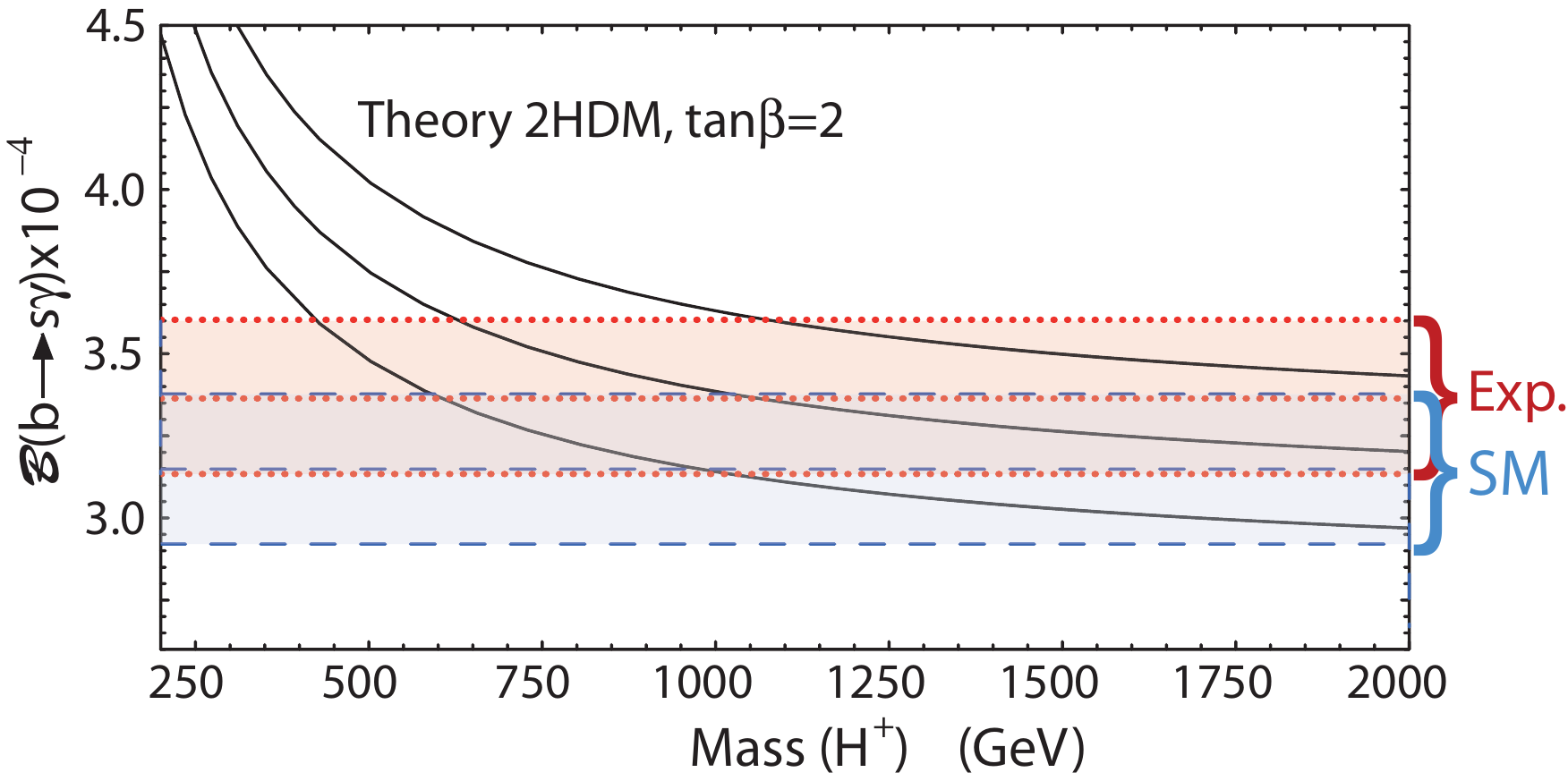} 
\caption{${\cal{B}}\left(b\to s\gamma\right)$ as a function of the charged Higgs boson mass. There are three sets of curves including the experimental average shown in dotted (red), the SM prediction in dashed (blue) and the 2HDM model in solid (black). The middle curve of each set gives the central value and higher and lower ones the $\pm 1\sigma$ bands \cite{Misiak-fig}. }\label{bsgamma-MHc-newBaBar}
\end{center}
\end{figure}

My average of measured branching fractions, including a recent result at this conference from BaBar \cite{Eigen}, is
\begin{equation}
{\cal{B}}\left(b\to s\gamma\right)=(3.37\pm 0.23)\times 10^{-4}~.
\end{equation}
The SM prediction is $(3.15\pm 0.23)\times 10^{-4}$ \cite{Misiak-fig,Misiak2}; thus the ratio of the measured to theoretical prediction is 1.07$\pm$0.10, which serves to severely limit many, if not most, NP models. An example of such a constraint for the two-Higgs-double model (2HDM), with the ratio of the vacuum expectation values of the two Higgs doublets having a value of $\tan\beta=2$, is given in Fig.~\ref{bsgamma-MHc-newBaBar}. The intersection of the lower edge of the NP prediction with the upper line of the experiment gives a limit at the approximately two standard deviation $(\sigma)$ level of $m_{H^+}>385$ GeV \cite{Newnum}. 

\begin{figure} [hbt]
\begin{center}
\includegraphics[width=1.\textwidth]{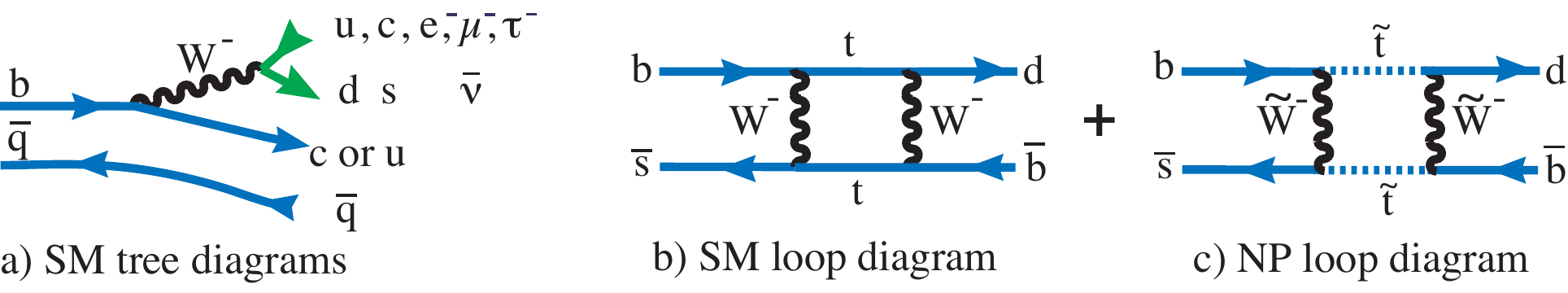} 
\caption{(a) Examples of SM tree-level diagrams. (b) Example of a SM loop diagram for $B_s$ mixing interfering that can interfere with (c) a NP diagram. }\label{tree-loop-diag}
\end{center}
\end{figure}

It is of course true that experiments measure the sum of SM and NP, so to set limits or see signals from NP we have to be confident in understanding the SM contribution. This was clear in the above example of $b\to s\gamma$. In other processes SM calculations are not possible, or are not unambiguous. An often used paradigm is that ``tree-level'' diagrams are dominated by SM processes, but loop level diagrams can be strongly influenced by NP. Examples of tree diagrams are given in Fig.~\ref{tree-loop-diag}(a), while both SM and loop diagrams that could have new as yet undiscovered particles in the loop are shown in Fig.~\ref{tree-loop-diag}(b); here both diagrams would add in amplitude and thus interfere allowing for either increases or decrease in rates and asymmetries.

NP can also be limited via a general study of loop processes, again assuming that it doesn't contribute to tree-level processes.
One can write a general local Lagrangian which includes an infinite tower of operators
with dimension $d > 4$, constructed in terms of SM fields, suppressed by inverse powers of an effective
scale $\Lambda>M_W$ as
\begin{equation}
\label{eq:Nphysops}
{\cal{L}}_{\rm eff} ={\cal{L}}_{\rm SM} +\sum{\frac{c_i^{(d)}}{\Lambda^{(d-4)}}{\cal{O}}^{(d)}_i},
 \end{equation}
where the ${\cal{O}}$'s represent the SM fields \cite{INP}. Lower bounds on $\Lambda$ are shown in Table~\ref{tab:ULscale}, taking the $c_i$ coefficients equal to one for different operators.

\begin{table} [htb]
\centering
\begin{tabular}{lcl} 
\hline
Operator & $\Lambda$ lower bound (TeV) & Observable\\\hline
Re$[(\overline{s}_L\gamma^{\mu}d_L)^2]$ & $9.8\times 10^2$ & $K_L-K_S$ mass difference, $\Delta m_K$\\
Im$[(\overline{s}_L\gamma^{\mu}d_L)^2]$ & $1.6\times 10^4$ &  $CP$ violation in $K_L$ decay, $\epsilon_K$\\
Re$[(\overline{s}_R d_L)(\overline{s}_L d_R)]$ & $1.8\times 10^4$ & $K_L-K_S$ mass difference, $\Delta m_K$\\
Im$[(\overline{s}_R d_L)(\overline{s}_L d_R)]$ & $3.2\times 10^5$ & $CP$ violation in $K_L$ decay, $\epsilon_K$\\
Re$[(\overline{b}_L\gamma^{\mu}d_L)^2]$ & $9.8\times 10^2$ & $B_L-B_S$ mass difference, $\Delta m_B$\\
Im$[(\overline{b}_L\gamma^{\mu}d_L)^2]$ & $1.6\times 10^4$ &  $CP$ violation in $B^0\to J/\psi K_S$ decay\\
Re$[(\overline{b}_R d_L)(\overline{s}_L d_R)]$ & $1.8\times 10^4$ & $B_L-B_S$ mass difference, $\Delta m_B$\\
Im$[(\overline{b}_R d_L)(\overline{s}_L d_R)]$ & $3.2\times 10^5$ & $CP$ violation in $B^0\to J/\psi K_S$ decay\\
\hline
\end{tabular}
\caption{Lower limits on the size of representative dimension six loop diagram operators ($\Lambda$), taking the coefficients $c_i$
equal to one. Adopted from Ref.~\cite{INP}.}
\label{tab:ULscale} 
\end{table}

These limits are impressive ranging from approximately 100 to 10,000 TeV, well beyond the direct collider limit search possibilities. Of course it is possible that the $c_i$ are smaller than one, but they have to be in range from $10^{-6}$ to $10^{-9}$ to bring these limits to the TeV scale. There are ways, however,  that NP can be consistent with these limits. One way is for the new particles in the loop diagrams to be degenerate in mass and have opposite phase, so they cancel in the loops. Another is for the quark mixing angles to be the same in the NP sector as in the SM. This defacto prescription, called ``Minimum Flavor Violation,'' is well described by Buras~\cite{MFV}. Of course very heavy new particles are also a possibility. Overall, these measurements put very severe constraints on NP models.

\section{\boldmath Mixing and $CP$ violation}
\begin{figure} [hbt]
\begin{center}
\includegraphics[width=.9\textwidth]{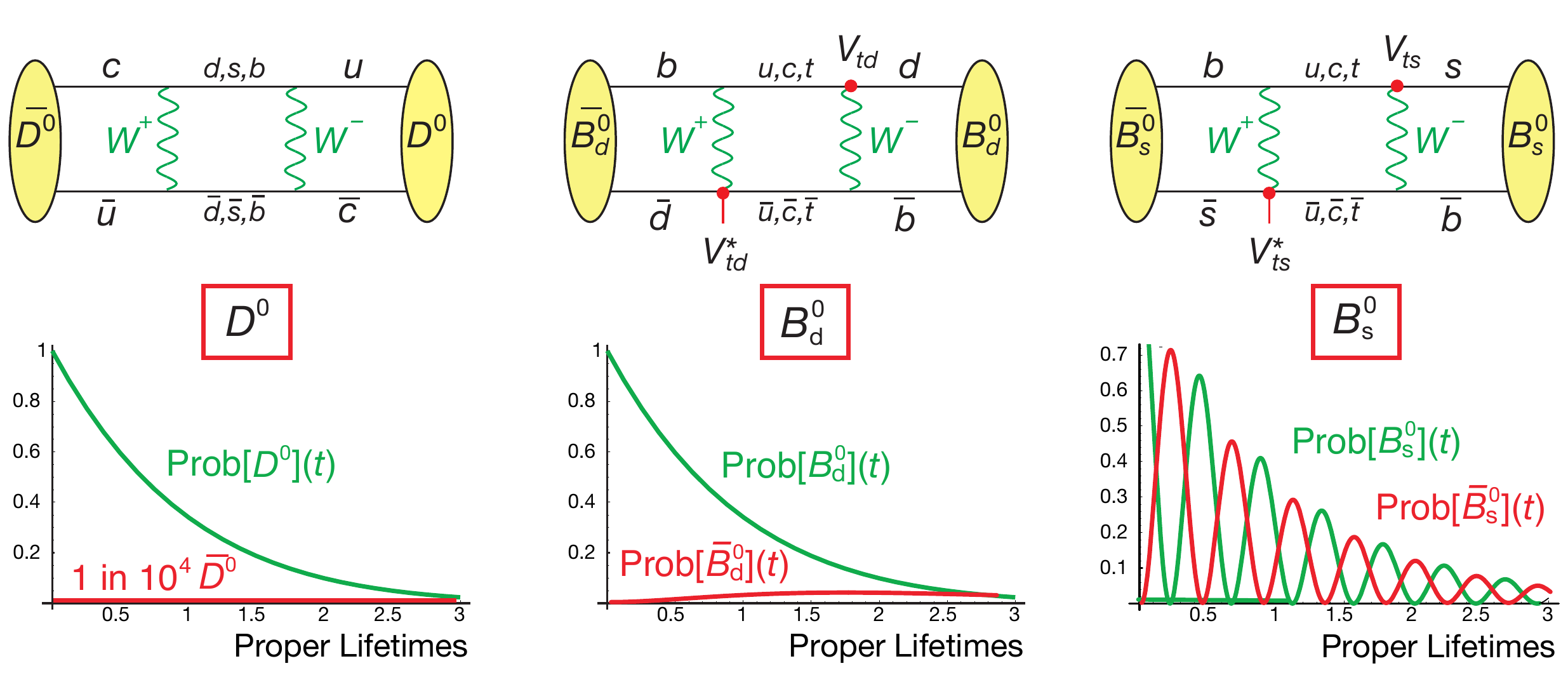} 
\caption{(top) Diagrams for $D^0$, $B^0$ and $B_s^0$ mixing. (bottom) Sketches of the time dependence for mesons to decay unmixed and mixed. (Adapted from Van-kooton \cite{Van-kooton}.)}\label{mix-rick}
\end{center}
\end{figure}

Let us consider in more detail the mixing diagrams which allow transformation of neutral mesons into their anti-particles. Fig.~\ref{mix-rick} shows the second order weak interaction diagrams for $D^0$, $B^0$ and $B_s^0$ mesons. The amplitude strength depends on the values of the CKM elements at the four vertices and approximately as the magnitude of the mass-squared of the virtual quarks permitted. $B^0$ and $B_s^0$ mixing transitions are far larger than that of the $D^0$, because the $t$ quark is allowed in the $B$ cases, but for the $D^0$ the heaviest allowed exchange quark is the $b$. Hence $D^0$ mixing is quite small, but finite \cite{HFAG}, while the $B^0$ is quite significant and the $B_s^0$ huge, the difference in the two $B$ cases is due entirely to relative sizes of the CKM elements. 

The Schr\"odinger equation is used to describe time dependent transformations between meson and anti-meson states for diagrams similar to the one in Fig.~\ref{tree-loop-diag}(b) in terms of a mass matrix $M$ and a decay matrix $\Gamma$:
\begin{equation}
i\frac{d}{dt}\left(\begin{array} {c} B^0_s \\ \overline{B}^0_s \end{array}\right)=
\left(\begin{array}{cc}M_{11}-i\Gamma_{11}/2  & M_{12}-i\Gamma_{12}/2 \\
M_{12}^*-i\Gamma_{12}^*/2  & M_{22}-i\Gamma_{22}/2 \end{array}\right)
\left(\begin{array} {c} B^0_s \\ \overline{B}^0_s \end{array}\right)
\label{eq:twostate}
\end{equation}
Diagonalizing we find $\ket{M_L}=p\ket{B^0_s}+q\ket{\overline{B}^0_s}$, $\ket{M_H}=p\ket{B^0_s}-q\ket{\overline{B}^0_s}$, where $p$ and $q$ are complex numbers that are equal to $1/\sqrt{2}$ in the absence of $CP$ violation. Also, assuming $CPT$ invariance, the mass $m(B_s)=M_{11}=M_{22}=(M_H+M_L)/2$, and the mass difference is defined as $\Delta M=M_H-M_L$, while for the decay widths, $\Gamma=1/\tau(B_s)=(\Gamma_H+\Gamma_L)/2=\Gamma_{11}=\Gamma_{22}$, and $\Delta\Gamma=\Gamma_L-\Gamma_H$. For convenience the variable $y\equiv\Delta\Gamma/2\Gamma$ is defined.

Time dependent $CP$ violation depends on the process in many ways.  If the final state $f$ is a $CP$ eigenstate then $f=\bar{f}$ and the $CP$ violating asymmetry is defined as \cite{Nierste}
\begin{equation}
\label{eq:CPviamix}
a(t)=\frac{\Gamma\left(\overline{M}\to f\right)-\Gamma\left({M}\to f\right)}{\Gamma\left(\overline{M}\to f\right)+\Gamma\left({M}\to f\right)}~.
\end{equation}
The amplitudes are defined as $A_f\equiv A(M\to f)$,  $\bar{A}_f\equiv \bar{A}(M\to f)$. The magnitude of the $CP$ violating effect for each channel depends on the CKM elements present in the decay, and is given by
\begin{equation}
\lambda_f=\frac{p}{q}\frac{\bar{A}_f}{A_f}~.
\end{equation}
The decay rates, when $|\lambda_f|=1$, are given by
\begin{eqnarray}
\Gamma\left(\overline{M}\to f\right)&=&N_f|A_f|^2e^{-\Gamma t} \left(\cosh\frac{\Delta\Gamma t}{2}-{\rm Re}\lambda_f\sinh\frac{\Delta\Gamma t}{2}+{\rm Im}\lambda_f\sin\left(\Delta M t\right)\right)\nonumber\\
\Gamma\left({M}\to f\right)&=&N_f|A_f|^2e^{-\Gamma t} \left(\cosh\frac{\Delta\Gamma t}{2}-{\rm Re}\lambda_f\sinh\frac{\Delta\Gamma t}{2}-{\rm Im}\lambda_f\sin\left(\Delta M t\right)\right)~.
\end{eqnarray}
Thus the $CP$ violating asymmetry is given by
\begin{equation}
\label{eq:abyt}
a(t)=\frac{{\rm Im}\lambda_f\sin\left(\Delta M t\right)}{\cosh\frac{\Delta\Gamma t}{2}-{\rm Re}\lambda_f\sinh\frac{\Delta\Gamma t}{2}}~.
\end{equation}

\subsection{\boldmath $CP$ violation in $\overline{B}^0_s$ decays}
\begin{figure} [hbt]
\begin{center}
\includegraphics[width=.6\textwidth]{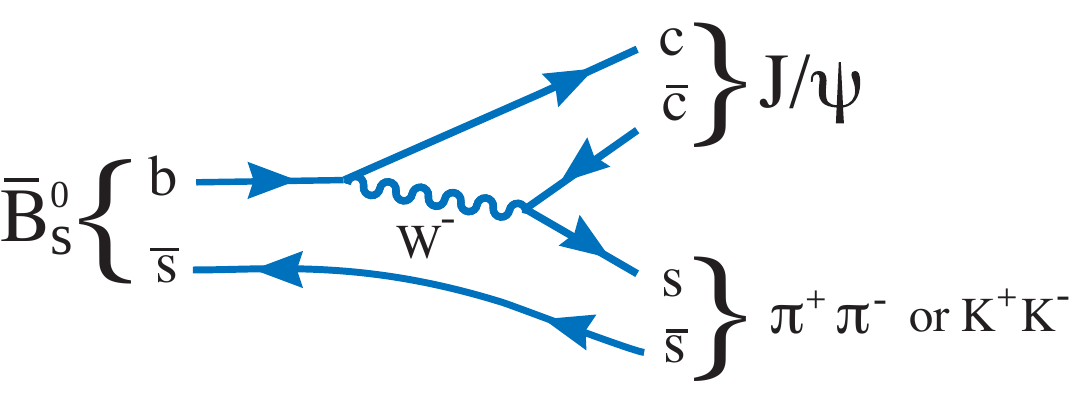} 
\caption{Tree level diagram for $\overline{B}^0_s\to J/\psi \pi^+\pi^-$ or $K^+K^-$ decays.}\label{BstoJpsipipi-KK}
\end{center}
\end{figure}

Let us consider the decays $\overline{B}^0_s\to J/\psi \phi$, where $\phi\to K^+K^-$ and $J/\psi\pi^+\pi^-$. The tree-level decay diagram is shown in Fig.~\ref{BstoJpsipipi-KK}. Here we use the interference between the direct decay and the one that proceeds via mixing: $\overline{B}^0_s\to B^0_s\to J/\psi \phi$, for example.
The $CP$ violating asymmetry $a(t)$ can be expressed in terms of the phase angle $\phi_s$, and the $CP$ parity of the final state, $\eta_{\rm CP}$.  In the SM ${\rm Im}(\lambda_f)= -\eta_{\rm CP}\sin\phi_s=\eta_{\rm CP}\arg\left(\frac{V_{ts}V^*_{tb}}{V^*_{ts}V_{tb}}
\frac{V^*_{cs}V_{cb}}{V_{cs}V^*_{cb}}\right)=-0.04$~rad for $CP$ odd states \cite{SM-pred}. 

\begin{figure}[htb]
 \begin{center}
   \includegraphics[width=3.7in]{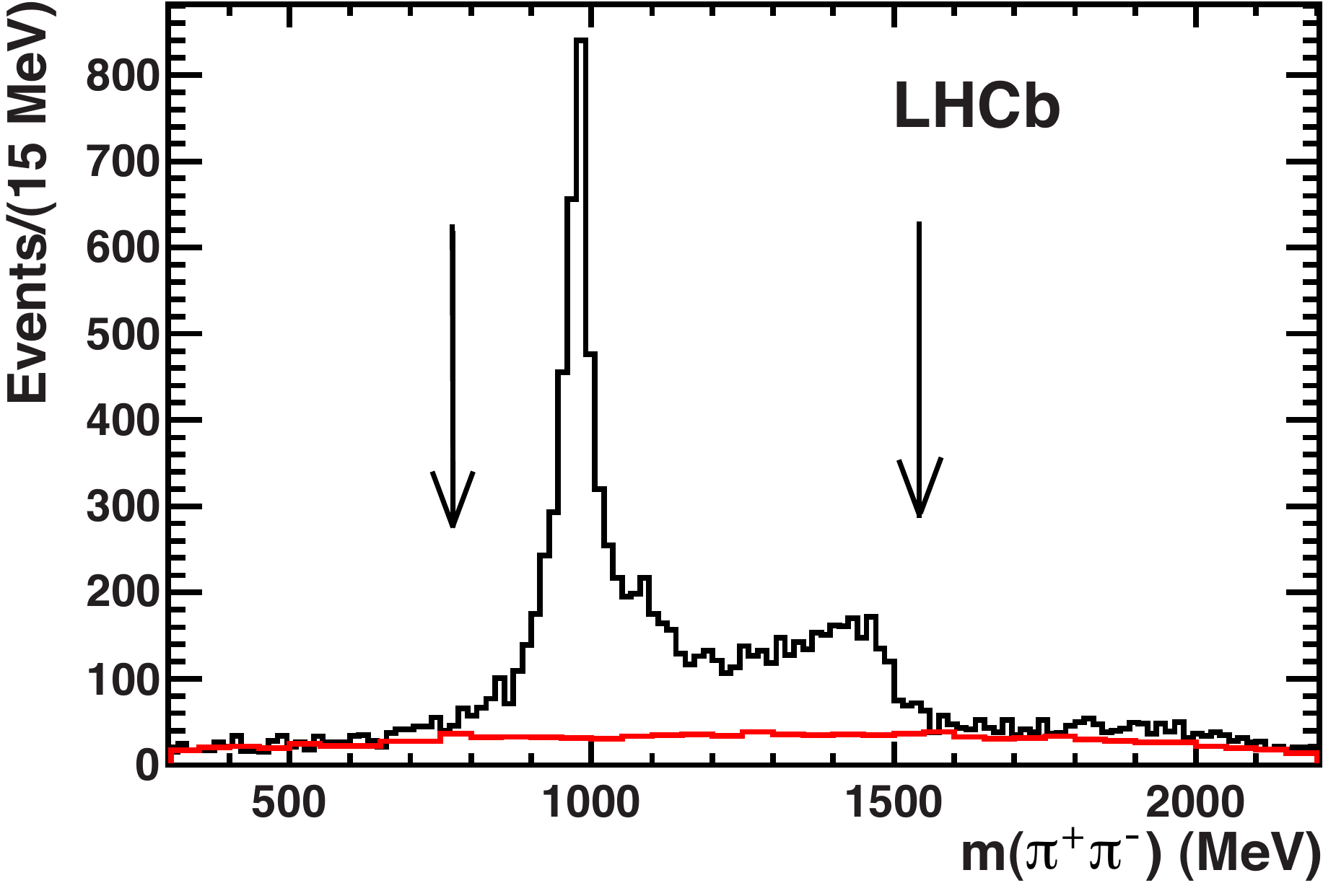}
   \caption {The $\pi^+\pi^-$ invariant mass in $\overline{B}^0_s\to J/\psi \pi^+\pi^-$ decays. The dashed (red) curve shows the background, and the arrows indicated the region chosen for $CP$ studies.
      }
   \label{mpipi-BDT}
   \end{center}
   \end{figure}

Based on a conjecture by Stone and Zhang \cite{Stone-Zhang}, the LHCb collaboration found the decay mode $\overline{B_s^0}\to J/\psi f_0(980)$ where $f_0(980)\to\pi^+\pi^-$,  close to the predicted level \cite{Jpsif0}, and was soon confirmed by others \cite{Jpsif02}. As the final state consists of a scalar $f_0(980)$ and a vector $J/\psi$, the relative angular momentum between the $f_0(980)$ and the $J/\psi$ is a P-wave and the final state is pure $CP$ odd. LHCb then examined the entire $\pi^+\pi^-$ mass spectrum in the $\overline{B}^0_s\to J/\psi \pi^+\pi^-$ channel, shown in Fig.~\ref{mpipi-BDT}.
The region between the arrows has a large signal to background ratio. In this region a full Dalitiz like analysis was performed that identified the individual components, and more importantly showed that the final state is $>97.7$\% CP-odd \cite{LHCb-jpsipipi-Daltiz}. Measurement of $CP$ violation based on Eq.~\ref{eq:abyt} gives a value 
$\phi_s=-0.019^{+0.173+0.004}_{-0.174-0.003}$~rad \cite{LHCb-jpsipipi-CP}.

\begin{figure}[htb]
 \begin{center}
   \includegraphics[width=3.5in]{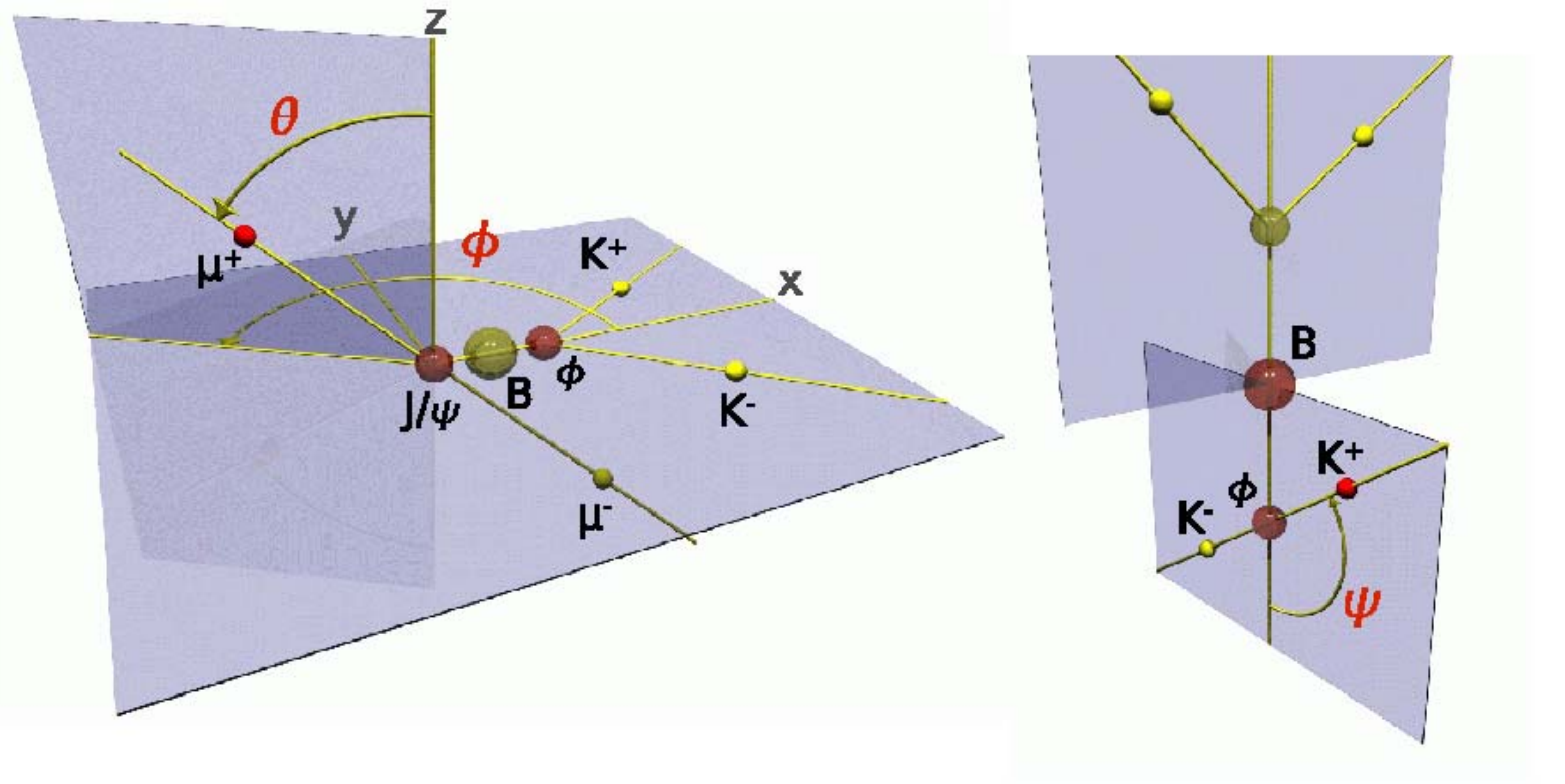}
   \caption{
    Pictoral description of the decay angles. On the left $\theta$ and $\phi$ defined
    in the $J/\psi$ rest frame and on the right $\psi$ defined in the
    $\phi$ rest frame. (From T. Kuhr \cite{trans-fig}.).
   }
   \label{transversity}
   \end{center}
   \end{figure}

The $J/\psi\phi$ state being composed of two vector particles is not a $CP$ eigenstate. Such final states still can be used but an angular analysis is necessary to separate the $CP$-even and $CP$-odd components \cite{transversitybib}, which requires fitting time dependent angular distributions. The three angles are shown in Fig.~\ref{transversity}.

In addition to the three \ P-wave amplitudes in the decay width, an additional amplitude due to possible $K^+K^-$ S-wave also must be considered \cite{Stone-Zhang}. The complexity of the analysis scheme is shown by writing the distribution of the signal decay time and angles, described by a
sum of ten terms, corresponding to the four polarization amplitudes
and their interference terms. Each of these is the product of a
time-dependent function and an angular function~\cite{transversitybib}
\begin{equation}
  \frac{\ud^{4} \Gamma({B}_s^0) }{\ud t \;\ud\theta \ud\psi \ud\phi} \; \propto \;
  \sum^{10}_{k=1} \: h_k(t) \: f_k( \theta,\psi, \phi) \,.
  \label{Eqbsrate}
\end{equation}
The time-dependent functions $h_k(t)$ can be written as
\newcommand{\coefcosh}{\ensuremath{a_k}} \newcommand{\coefsinh}{\ensuremath{b_k}}
\newcommand{\coefcos}{\ensuremath{c_k}} \newcommand{\coefsin}{\ensuremath{d_k}}
\begin{multline}
  h_k (t) \; = \; N_k e^{- \Gs
    t} \: \left[ 
    \coefcos \cos(\dms t) \, 
    + \coefsin \sin(\dms t)
    \,  
    + \coefcosh \cosh\left(\tfrac{1}{2} \DGs t\right) 
    + \coefsinh \sinh\left( \tfrac{1}{2} \DGs t\right)
  \right] . 
  \label{equ:timefunc}
\end{multline}
where $\dms{}$ is the $B^0_s$ oscillation frequency. The coefficients
$N_k$ and $a_k,\ldots,d_k$ can be expressed in terms of $\phi_s$ and
four complex transversity amplitudes $A_i$ at $t=0$. The label $i$
takes the values $\{\perp,\parallel,0\}$ for the three P-wave
amplitudes and S for the S-wave amplitude. For a particle produced in a $B^0_s$ flavour eigenstate the
coefficients in Eq.~\ref{equ:timefunc} and the angular functions
$f_k(\theta,\psi, \phi)$ are then, given in Table~\ref{tab:Omega}
\cite{ Xie}, where $\delta_0=0$ is chosen arbitrarily
\newcommand{\cosphis}{\cos\phi_s}
\newcommand{\sinphis}{\sin\phi_s}

\begin{table}[htb]
\label{tab:Omega}
\begin{center}
\vspace{-6mm}
\begin{equation}{\small
\begin{array}{c|c|c|c|c|c|c}
\hline\hline
  k  & f_k(\theta,\psi, \phi) & N_k                & \coefcosh                  & \coefsinh & \coefcos & \coefsin \\
    \hline
  1  & 2\,\cos^2\psi \left(1 - \sin^2\theta \cos^2\phi\right) & |A_0(0)|^2         & 1                          & -\cosphis & 0 & \sinphis \\
  2  & \sin^2\psi \left(1 - \sin^2\theta \sin^2\phi\right)    & |A_\|(0)|^2         & 1                          & -\cosphis & 0 & \sinphis \\
  3  & \sin^2\psi \sin^2\theta                                & |A_\perp(0)|^2      & 1                          & \cosphis & 0 & -\sinphis  \\
  4  & -\sin^2\psi \sin2\theta \sin\phi                      & |A_\|(0)A_\perp(0)| & 0                          & -\cos(\delta_{\perp {\|}})\sinphis & \sin(\delta_{\perp {\|}}) & -\cos(\delta_{\perp {\|}})\cosphis  \\
  5  & \tfrac{1}{2}\sqrt{2} \sin2\psi \sin^2\theta \sin2\phi & |A_0(0) A_\|(0)|   & \cos(\delta_{\| 0})   & -\cos(\delta_{\| 0}) \cosphis & 0 & \cos(\delta_{\| 0}) \sinphis \\
  6  & \tfrac{1}{2}\sqrt{2} \sin2\psi \sin2\theta \cos\phi   & |A_0(0) A_\perp(0)| & 0                          & -\cos(\delta_{\perp 0}) \sinphis  & \sin(\delta_{\perp 0}) & - \cos(\delta_{\perp 0}) \cosphis \\
  7  & \tfrac{2}{3} (1-\sin^2\theta\cos^2\phi)              & |A_\rmS(0)|^2         & 1                          & \cosphis & 0 & -\sinphis  \\
  8  & \tfrac{1}{3}\sqrt{6}\sin\psi\sin^2\theta\sin 2\phi   & |A_\rmS(0) A_\|(0)|   & 0                          & -\sin(\delta_{\| \mathrm{S}}) \sinphis  & \cos(\delta_{\| \mathrm{S}}) & - \sin(\delta_{\| \mathrm{S}}) \cosphis \\
  9  & \tfrac{1}{3}\sqrt{6}\sin\psi\sin2\theta\cos\phi & |A_\rmS(0) A_\perp(0)| & \sin(\delta_{\perp\mathrm{S}}) & \sin(\delta_{\perp\mathrm{S}}) \cosphis & 0 & -\sin(\delta_{\perp\mathrm{S}})\sinphis \\
  10 & \tfrac{4}{3}\sqrt{3}\cos\psi(1-\sin^2\theta\cos^2\phi) & |A_\rmS(0) A_0(0)|    & 0                          & -\sin(\delta_{0\mathrm{S} })\sinphis & \cos(\delta_{0\mathrm{S} }) &  -\sin(\delta_{0\mathrm{S} })\cosphis  \nonumber\\
  \hline\hline
\end{array}
}\end{equation}
  \caption{The components of the decay width contributing to Eqs.~3.6 and 3.7,
where $\delta_{ij}\equiv \delta_i-\delta_j$, e.g. $\delta_{\perp 0}=\delperp- \delzero$.}

\end{center}
\end{table}

The differential decay rates for a $\overline{B}^0_s$ meson produced at time $t=0$
are obtained by changing the signs of $\phi_s$, $A_{\perp}(0)$ and
$A_\rmS(0)$, or, equivalently, the signs of \coefcos{} and \coefsin{}.\footnote{The decay width is invariant under the transformation
$(\phi_s,\DGs,\delpar,\delperp,\delta_\rmS) \mapsto
(\pi-\phi_s,-\DGs,-\delpar,\pi-\delperp,-\delta_\rmS)$ which gives
rise to a two-fold ambiguity in the results. This ambiguity has been removed by analyzing the interference between the S and P waves \cite{ambiguity}.} Note, that for the $J/\psi\pi^+\pi^-$ final state, only line \#7 in Table~\ref{tab:Omega} contributes.

\begin{figure} [hbt]
\begin{center}
\includegraphics[width=.7\textwidth]{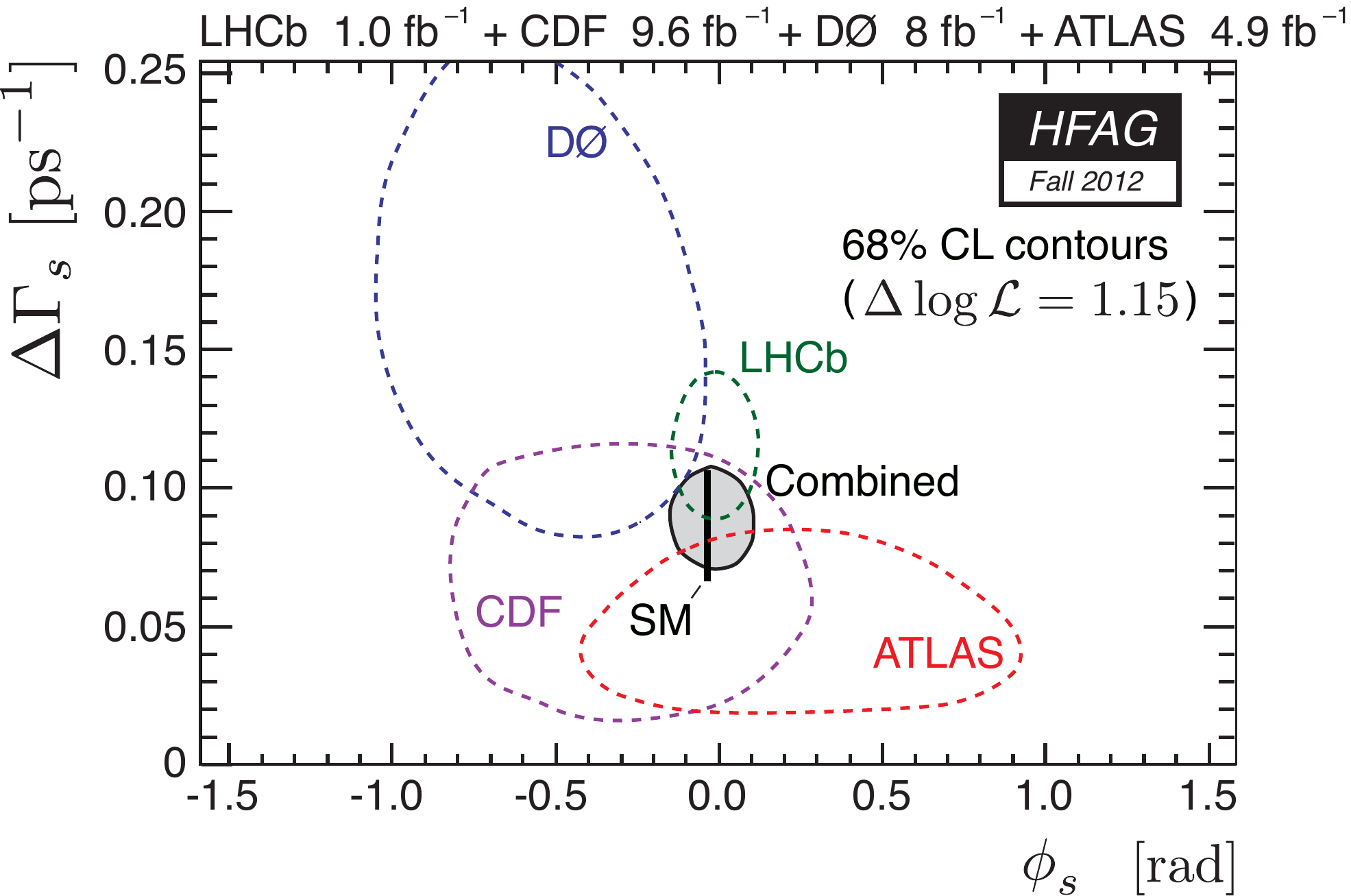} 
\caption{Measurements of $\Delta\Gamma$ versus $\phi_s$ as compiled by HFAG \cite{HFAG} .
} \label{phis-DG_comb} 
\end{center}
\end{figure}

Results of the analyses are presented in the $\Delta\Gamma_s - \phi_s$ plane, since earlier low statistics data had significant correlations. The current results are shown in Fig.~\ref{phis-DG_comb} for several experiments. The LHCb result, $\phi_s=0.001\pm 0.101\pm0.027$~rad is the most accurate and dominates the average. Combining with the $J/\psi\pi^+\pi^-$ result LHCb finds \cite{LHCb-Jpsiphi}
\begin{equation}
\phi_s=-0.002\pm0.083\pm0.027~{\rm rad,}
\end{equation}
which is consistent with the SM prediction.

\begin{figure} [hbt]
\begin{center}
\includegraphics[width=.7\textwidth]{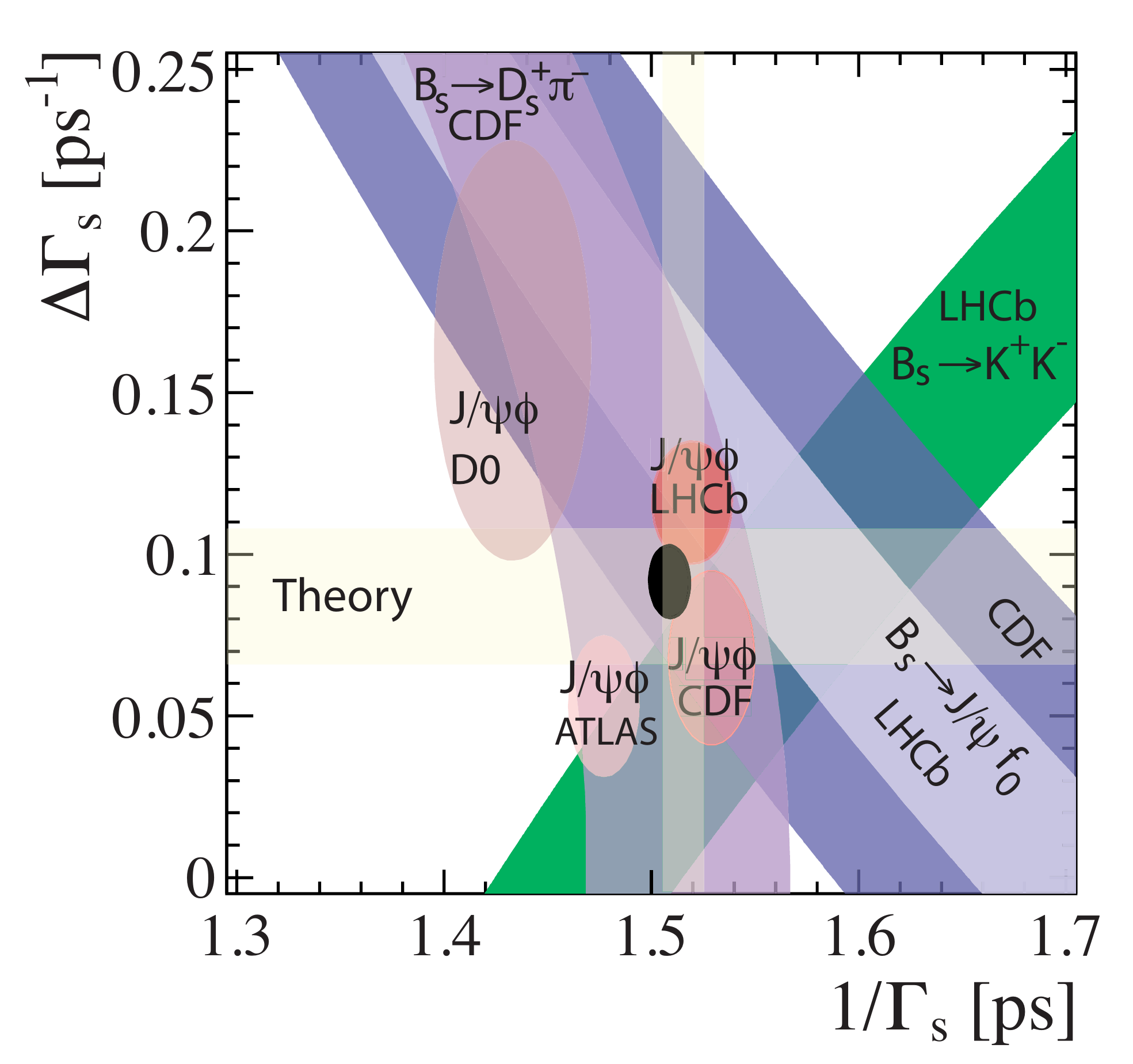} 
\caption{Measurements of $\Delta\Gamma_s$ versus the effective lifetime in each mode labeled as $1/\Gamma_s$, and the average as shown in the black ellipse as determined by A. Phan \cite{Anna}. The ovals show 39\% CL contours, while the bands are at 68\% CL. 
} \label{atlas_d0_cdfds_lhcb_update-theory} 
\end{center}
\end{figure}

I choose to determine the $\overline{B}_s^0$ lifetime and  $\Delta\Gamma_s$ based on measurements of only fully reconstructed decays where all the final state particles are seen.  
Fig.~\ref{atlas_d0_cdfds_lhcb_update-theory} shows the data from four separate determinations using $J/\psi\phi$, two using $J/\psi f_0(980)$, one using $D_s^+\pi^-$ at fully mixed $CP$ state, and one using $K^+K^-$.The last mode is a $CP$-even eigenstate where the effects of $CP$ violation are taken to be negligible. The results to an overall fit are shown in the black oval covering 39\% confidence level (CL). They are
\begin{eqnarray}
\tau_s &=& 1.509\pm 0.010~{\rm ps},\nonumber \\
\Delta\Gamma_s &=&0.092\pm0.011~{\rm ps}^{-1},\nonumber \\
y_s&=&\frac{\Delta\Gamma_s}{2\Gamma_s}= 0.07\pm0.01~.
\end{eqnarray}
These result agree very well with the theoretical predictions.

There is another way $CP$ violation can show itself in $\overline{B}^0$ decays. Consider the asymmetry defined as
\begin{equation}
\label{eq:CPmix}
a(t)=\frac{\Gamma\left(\overline{M}\to f\right)-\Gamma\left({M}\to \overline{f}\right)}{\Gamma\left(\overline{M}\to f\right)+\Gamma\left({M}\to \overline{f}\right)}~.
\end{equation}
The difference here with Eq.~\ref{eq:CPviamix} is that the final state $f$ is flavour specific, $f\ne \overline{f}$, and at zero decay time both $\Gamma\left(\overline{M}\to f\right)=0$ and $\Gamma\left({M}\to \overline{f}\right)=0$. For example, in the case of $\overline{B}^0_s$ decays the asymmetry between the decay rates for $\overline{B}^0_s\to D^+_s\mu^-\overline{\nu}$ and ${B}^0_s\to D^-_s\mu^+{\nu}$ can be measured. Another method is to use events where both $b$-flavoured hadrons decay semileptonically, i.e. to $X\mu\nu$. When one $b$ mixes and the other decays these events produce like-sign muons, and the difference between $\mu^+\mu^+$ and $\mu^-\mu^-$ events can be examined.

\begin{figure} [hbt]
\begin{center}
\vspace{-4mm}
\includegraphics[width=.5\textwidth]{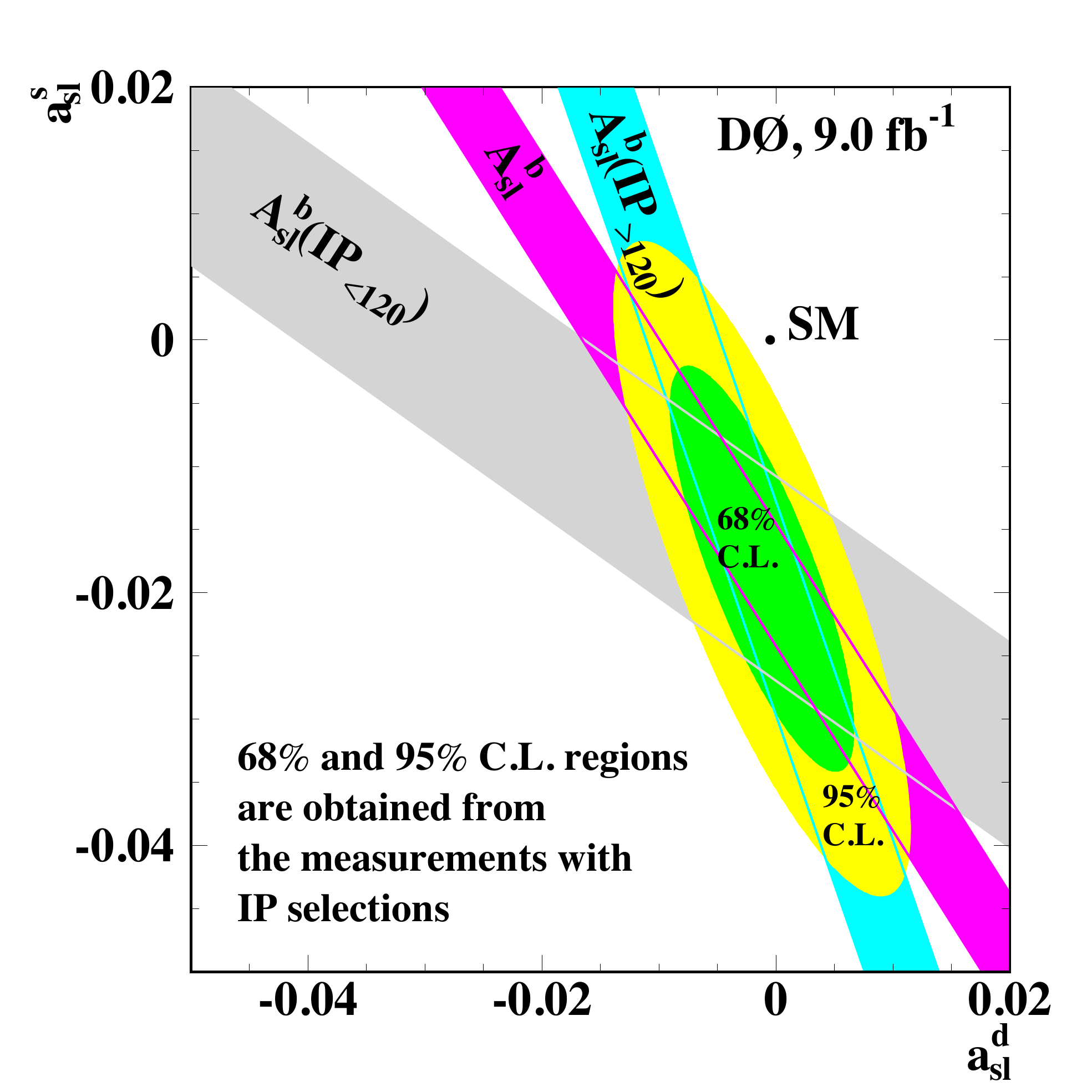} 
\vspace{-4mm}
\caption{Measurements of the semileptonic asymmetry using dimuon events from D0. The purple band ($A^b_{sl}$) is their base result, while the bands have impact parameter (IP) cuts applied that change the relative $B^0$ and $B_s^0$ yields \cite{Borrisov}.
} \label{d0-aslb-Borrisov} 
\end{center}
\end{figure}

In the SM the asymmetry is related to decay width as $a_{sl}=\left(\Delta\Gamma/\Delta M\right)\tan\phi_{12}$, where $\tan\phi_{12}=-\arg\left(-\Gamma_{12}/M_{12}\right)$. The asymmetries are expected to be very small in the SM, $-4.1\times 10^{-4}$, and $1.9\times 10^{-5}$ for $B^0$ ($a_{sl}^d$) and $B_s^0$ ($a_{sl}^s$), respectively \cite{Lenz}. The D0 collaboration, however, reported an anomalously large value for the dimuon asymmetry of $A^b_{sl}=(-0.787\pm 0.172 \pm 0.093)$\% \cite{D0-asls}. As they are summing over $B^0$ and $B_s^0$ decays, they graph their results as the purple band shown in Fig.~\ref{d0-aslb-Borrisov}. As the width of the band corresponds to $\pm 1\sigma$ uncertainty, their result is 3.9$\sigma$ from the SM value. If confirmed, this would be a clear indication of NP.

\begin{figure} [h!]
\vspace{-1mm}
\begin{center}
\includegraphics[width=.55\textwidth]{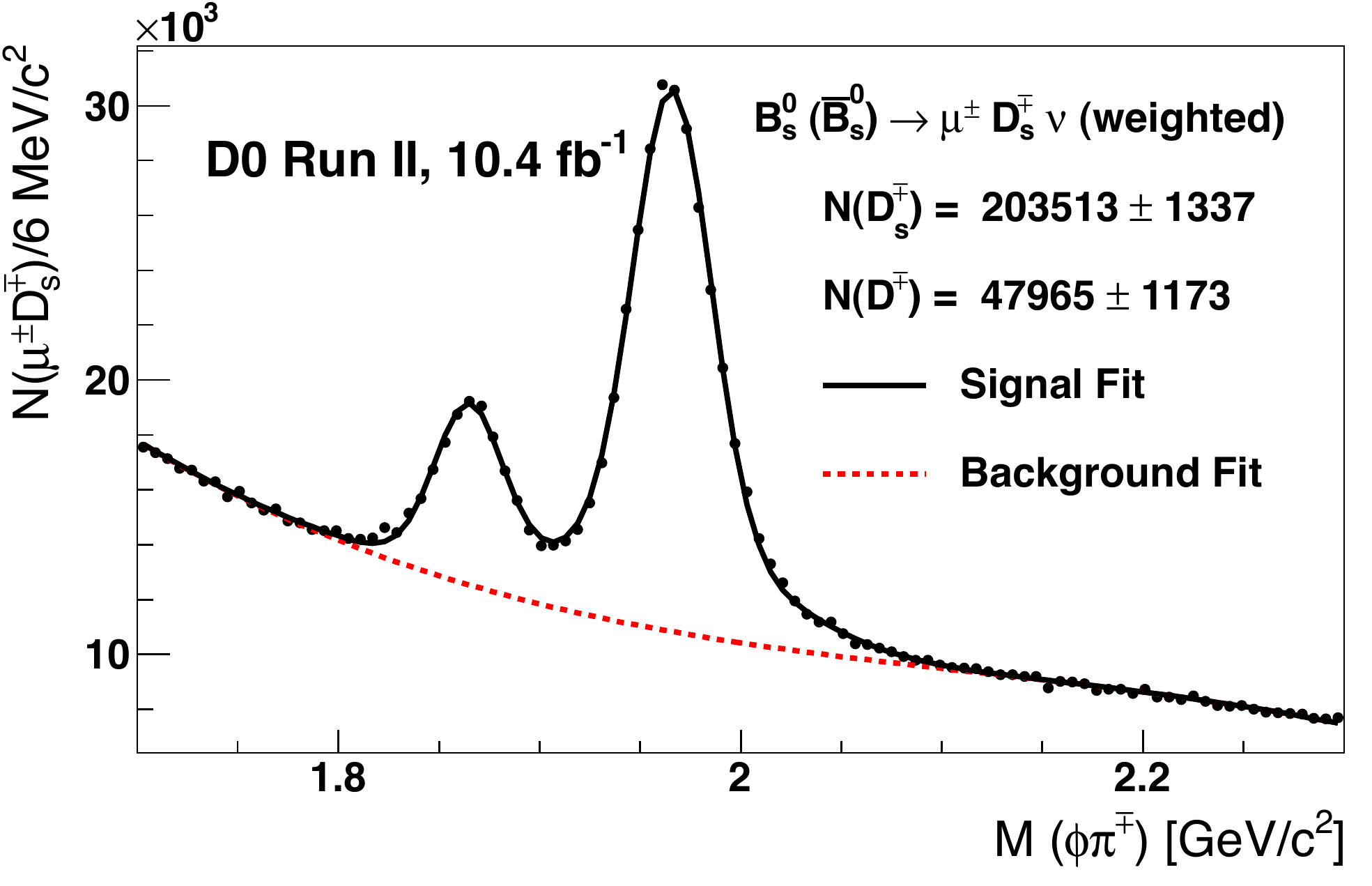} 
\vspace{-2mm}
\caption{The $\phi\pi^{\pm}$ mass spectrum for events where the tracks form a common vertex detached from the primary. The peaks are at the $D^+$ and $D_s^+$ masses. (Note the zero suppressed vertical scale.)
} \label{D0-Dsmunu} 
\end{center}
\end{figure}

Subsequently D0 measured the individual semleptonic asymmetries $a_{sl}^s$ 
\cite{D0_Dsmunu} and $a_{sl}^d$ \cite{D0-D+munu} using semileptonic $B$ decays, where a single muon is detected in conjunction with a charmed meson. The invariant $K^{+}K^{-}\pi^{\pm}$ mass distribution, where the $K^+K^-$ is required to be consistent with the $\phi$ meson mass,  for putative ${B}_s^0\to D_s^{\pm}\mu^{\mp}X$ decays is shown in Fig.~\ref{D0-Dsmunu}, where the $X$ indicates a missing neutrino plus possibly additional mesons.  The events in the $D_s$ signal peak come predominantly from $B_s^0$ decays. The measurement results from counting the number of $D_s^+\mu^-$ versus $D_s^-\mu^+$ events and correcting for any residual detector asymmetries. From these data D0 finds $a_{sl}^s=(-1.08\pm 0.72\pm 0.17)$\%, and from studying $D^{\pm}\mu^{\mp}X$ decays, with $D^{\pm}\to K^{\mp}\pi^+\pi^-$,  $a_{sl}^d=(0.68\pm 0.45\pm 0.14)\% $.

\begin{figure} [hbt]
\begin{center}
\vspace{-10mm}
\includegraphics[width=.9\textwidth]{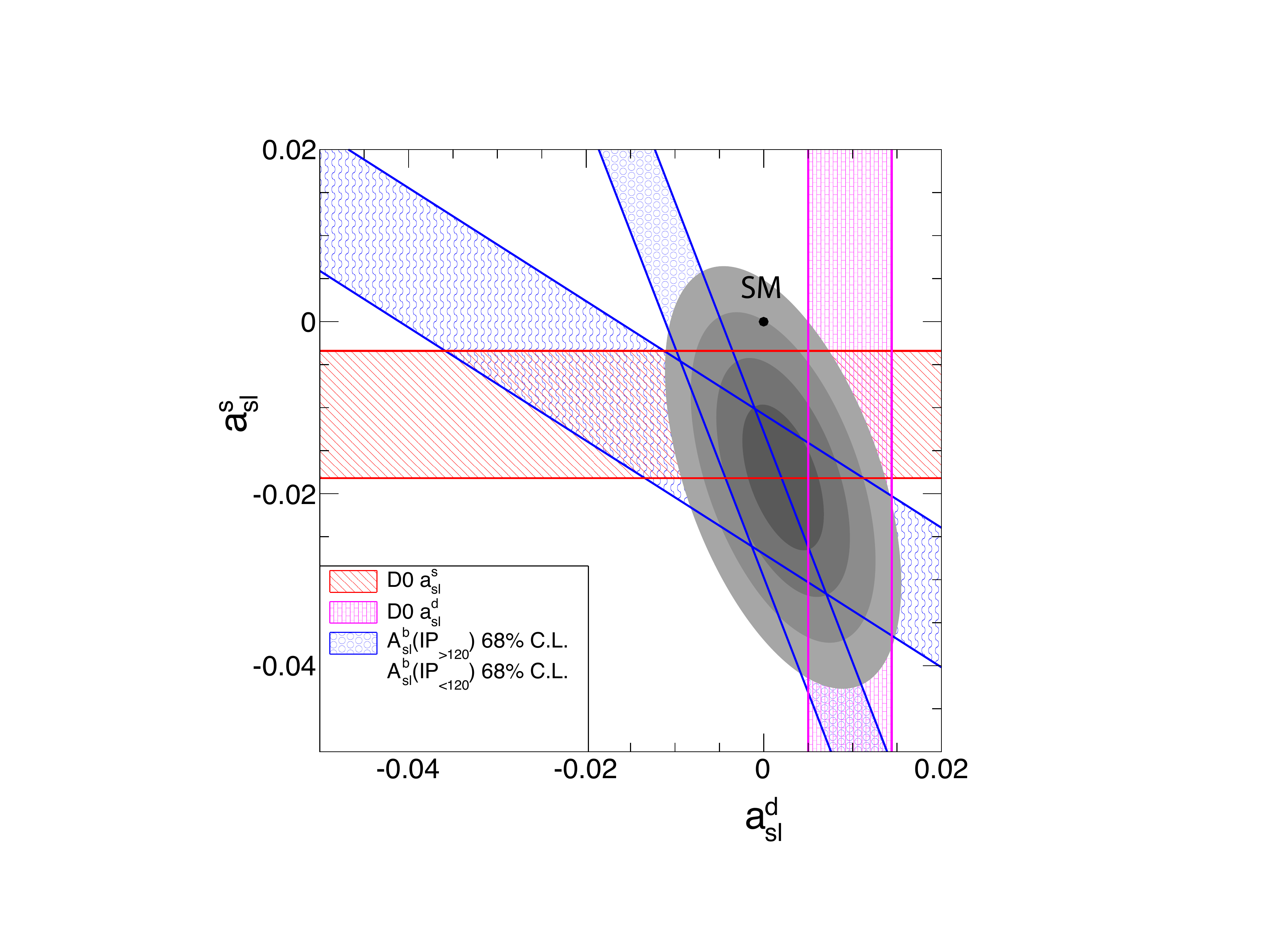} 
\vspace{-10mm}
\caption{Combination D0 semileptonic decay asymmetry measurements with
measurements using di-muons separated by into two impact parameter categories
by a cut of 120~$\mu$m. 
The error bands
represent $1\sigma$ uncertainties on each individual
measurement. The ellipses represent the 1, 2, 3, and
4$\sigma$ two-dimensional C.L. regions, respectively. The SM point is
shown with a black dot.} \label{D0-Bertram-sum} 
\end{center}
\end{figure}

Combining their di-muon results, including some separation into $a_{sl}^s$ and $a_{sl}^d$ values based on muon impact parameter, with the semileptonic measurements, Bertram compares the D0 measurements with the SM in Fig.~\ref{D0-Bertram-sum} \cite{Bertram}.

\begin{figure}[!h]
\begin{center}
\includegraphics[width= 5.8 in]{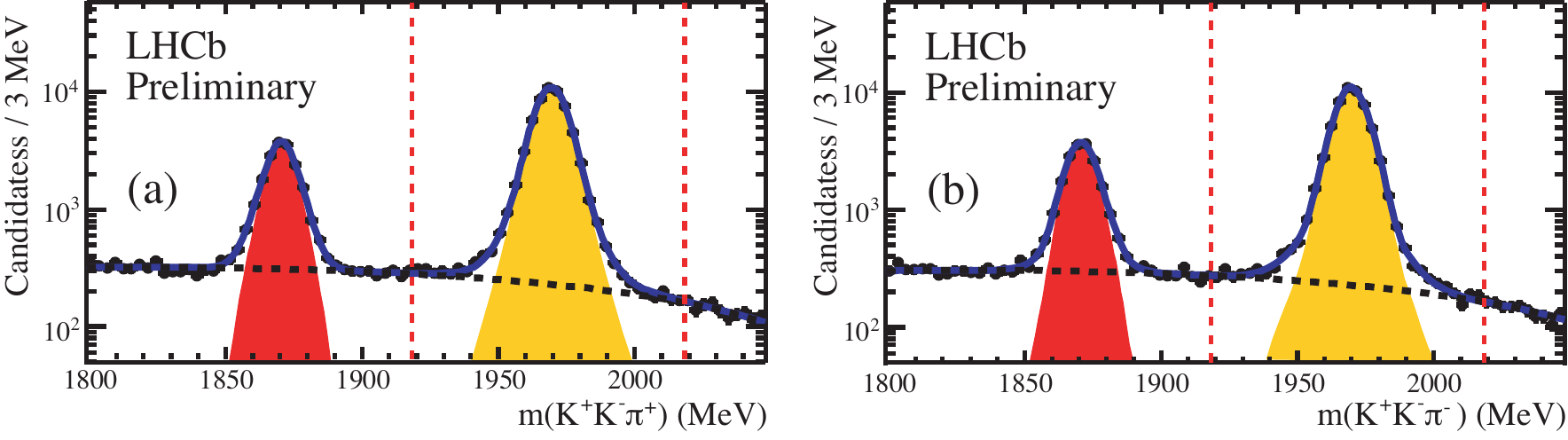}
\end{center}
\vspace{-5mm}
\caption{Invariant mass distributions for (a) $K^+K^-\pi^+$ candidates and (b)  $K^+K^-\pi^-$ candidates for magnet down data, requiring that $m(K^+K^-$) be  within $\pm$20~MeV of the $\phi$ meson mass. The vertical dashed lines indicate the signal region.} 
\label{LHCb-m_KKpi_-down}
\end{figure}

\begin{figure} [hbt]
\begin{center}
\includegraphics[width=.7\textwidth]{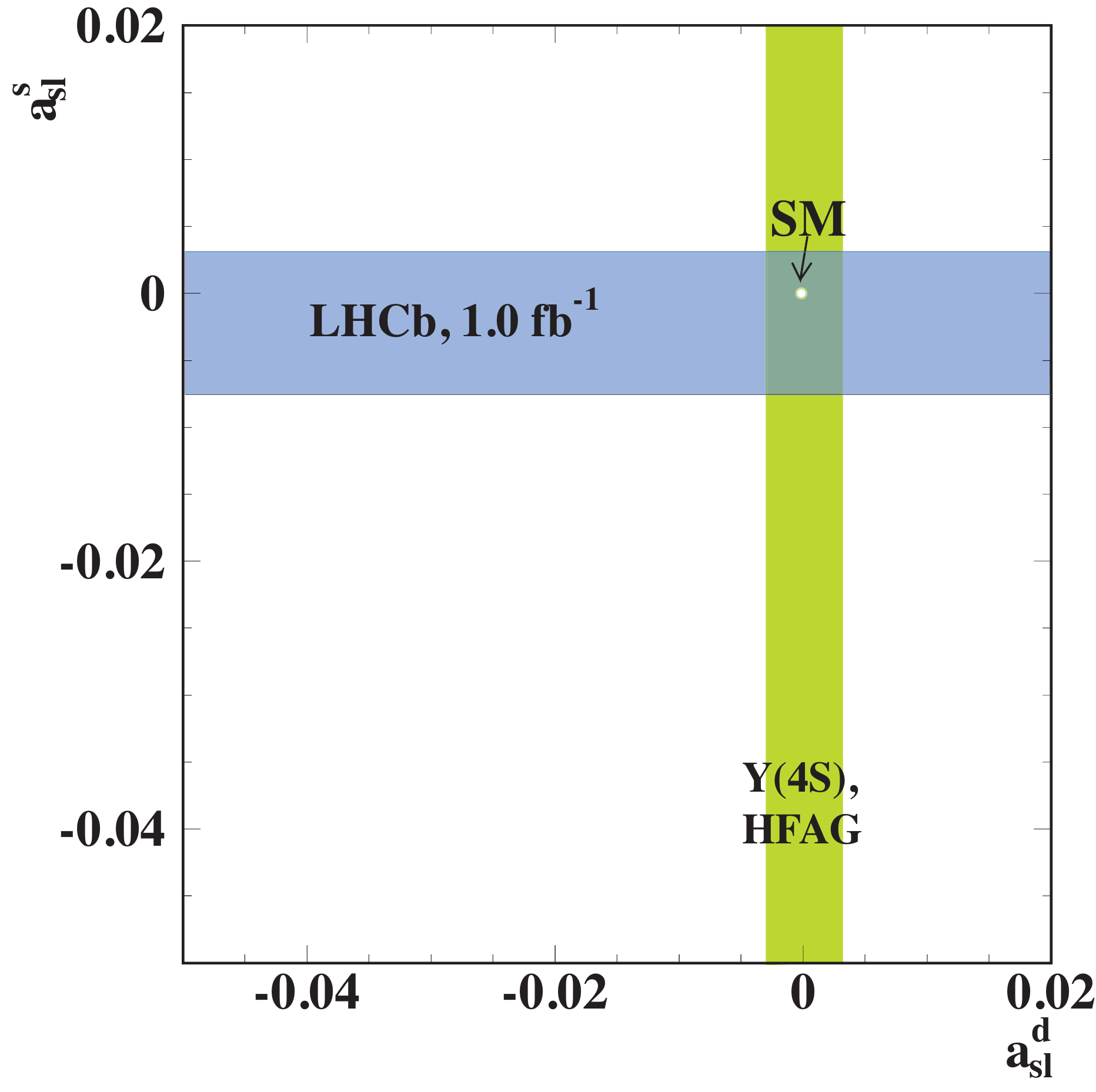} 
\caption{The LHCb measurement of $a_{sl}^s$, and the B-factory measurement of $a_{sl}^d$. The band edges are the central values  $\pm 1\sigma$ uncertainties. } \label{Asls-vs-Asld} 
\end{center}
\end{figure}

There is a differing world view on these asymmetries. The B-factories have produced measurements of $a_{sl}^d$ that average to  $0.0002\pm 0.0031$ \cite{HFAG}. LHCb recently measured the asymmetry in semileptonic $B_s^0$ decays, in a similar manner as D0, using a 1.0~fb$^{-1}$ data sample. The $K^+K^-\pi^{\pm}$ mass spectra is shown in Fig.~\ref{LHCb-m_KKpi_-down} for candidate decays with a muon of opposite charge to the candidate $D_s$ meson. LHCb periodically reverses the polarity of their spectrometer magnet, as did D0. This data is for one configuration only (magnetic field down).
The excellent mass resolution allows the $D^+$ and $D_s^+$ signals to be completely separated.
LHCb has different detection efficiencies, in principle, for $\pi^+$ versus $\pi^-$ and $\mu^+$ versus $\mu^-$. In each magnet setting the tracking efficiency differences mostly cancel considering the binary pairs $\pi^+\mu^-$ versus $\pi^-\mu^+$.  Most of the remaining differences are cancelled by averaging the two opposite magnet polarities.The different muon triggering efficiencies are evaluated using a sample of about one million $J/\psi\to\mu^+\mu^-$ found in events that were triggered independently of the $J/\psi$. LHCb finds $a_{sl}^s= (-0.24\pm0.54\pm0.33)$\% \cite{LHCb-CONF-2012-022}. Measurements from the B-factories and LHCb are plotted in Fig.~\ref{Asls-vs-Asld}. The data are completely consistent with the SM, and differ somewhat from the D0 results.

\afterpage{\clearpage}

\subsection{\boldmath Mixing and $CP$ violation in Charm}

LHCb recently measured mixing in the $D^0-\overline{D}^0$ system at more than 5$\sigma$ significance \cite{D0-mixing}, consistent with the average of several other less significant results \cite{HFAG}. Measurement of charm mixing at the levels seen is expected in the SM. Measurement of  $CP$ violation at the $\sim1$\% level, however, was not expected. An interesting paper discussed the possibility of seeing NP in charm $CP$ violation specifically by looking at singly Cabibbo suppressed decays \cite{GKN}.

Using the same definition for asymmetry as in Eq.~\ref{eq:CPviamix}, the Cabibbo suppressed final states $D^0\to K^+K^-$ and $D^0\to\pi^+\pi^-$ have been investigated. A flavor tag is provided by using $D^0$ that result from $D^{*\pm}\to\pi^{\pm}D^0$ decays. In order to cancel the effects of $D^{*\pm}$ production asymmetries and $\pi^+$ versus $\pi^-$ detection asymmetries, the difference between time-integrated asymmetries
$\Delta A_{CP}=a_{CP}(K^+K^-)-a_{CP}(\pi^+\pi^-)$ is determined. LHCb first published a 3.5 standard deviation result \cite{LHCbCharmCPV}. When averaging data from CDF \cite{CDFCharmCPV} and Belle \cite{BelleCharmCPV}, I find $\Delta A_{CP}=(-0.74\pm 0.15)$\%. However, none of the experiments by itself has a result of more than 3.5$\sigma$ significance, and more data are needed to see if the result is real. Meanwhile the theory community continues to argue if this asymmetry difference can be explained by the SM \cite{NPorSM}. 
I choose to treat this as a limit on NP of 1\%$>-\Delta A_{CP}>0$\%.

\section{Exclusive rare decays}

We have seen that the inclusive ``rare'' decay $b\to s\gamma$ has a relatively accurate SM prediction (see Sec.~\ref{sec:NP}). Other well determined predictions exist for exclusive rare decays. Since these decays have suppressed branching fractions the possibility that NP can make large contributions is enhanced. 

\subsection{\boldmath $\overline{B}_s^0\to\mu^+\mu^-$}

\begin{figure} [hbt]
\begin{center}
\vspace{-1mm}
\includegraphics[width=1.\textwidth]{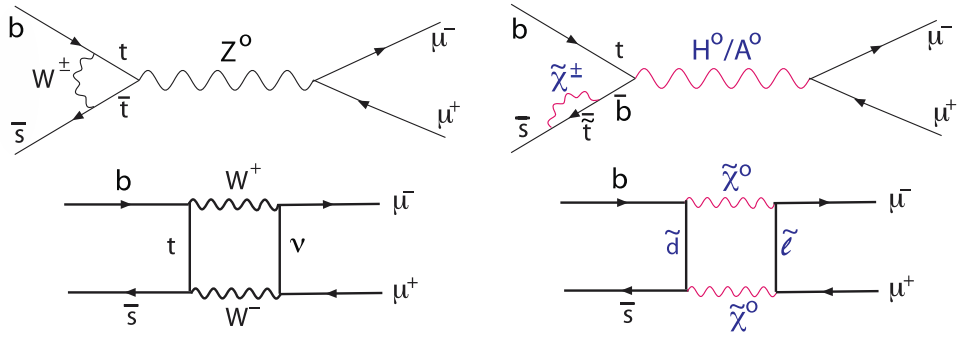} 
\vspace{-73mm}
\caption{Decay diagrams for $\overline{B}_s^0\to\mu^+\mu^-$ in the SM (left) and in
a particular NP model, that of the MSSM (right). } \label{Bs-mumu-diags} 
\end{center}
\end{figure}

\begin{figure} [hbt]
\begin{center}
\includegraphics[width=.6\textwidth]{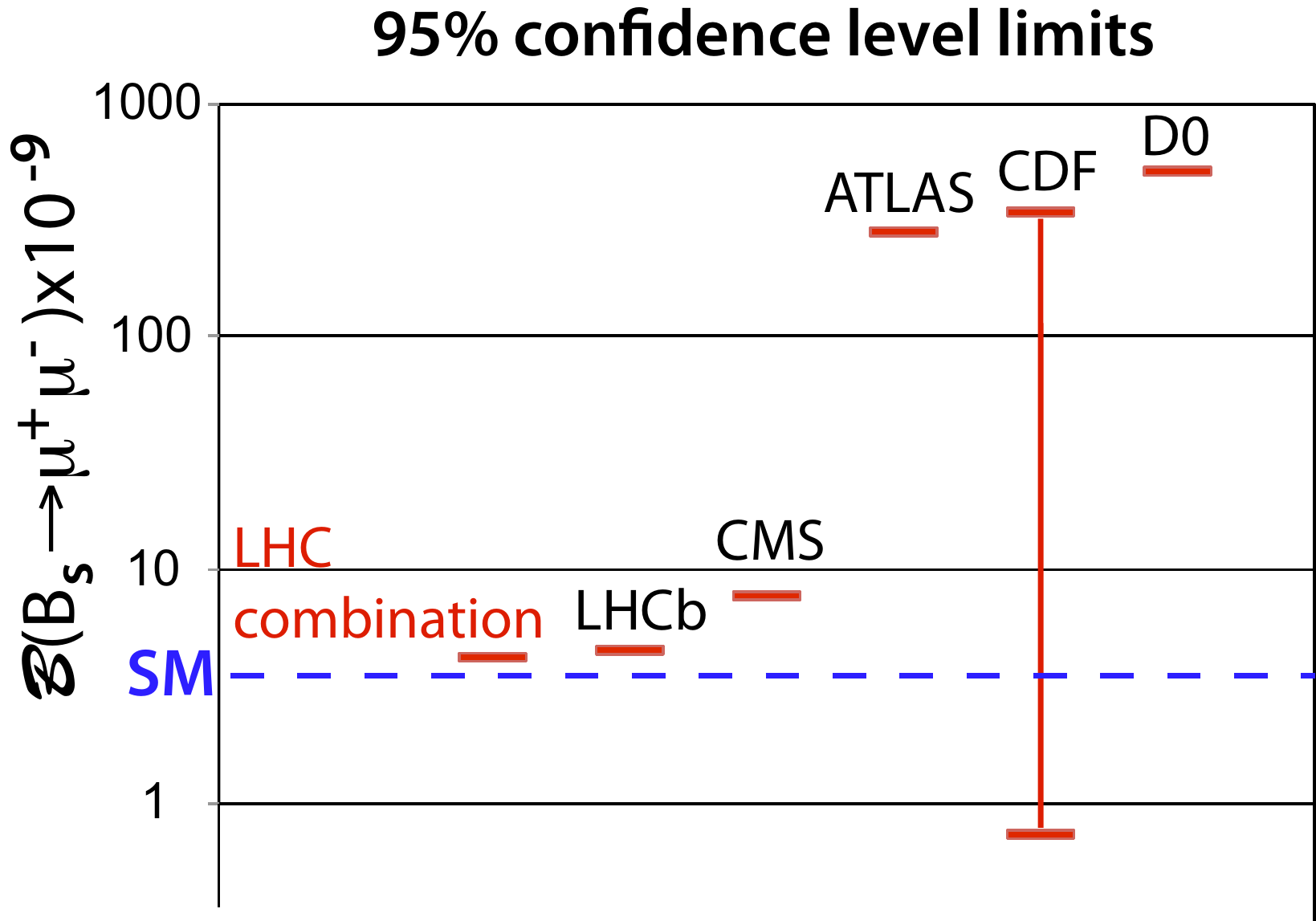} 
\caption{Upper limits for $\overline{B}_s^0\to\mu^+\mu^-$ from different experiments, as of July 4, 2012. The ``LHC combination," includes individual ATLAS, CMS and LHCb contributions shown in the figure.} \label{UL} 
\end{center}
\end{figure}

Heavily suppressed in the SM partially due to helicity conservation, the decay $\overline{B}_s^0\to\mu^+\mu^-$ is predicted by Buras \cite{Buras-Bsmumu} to have a branching fraction, after a correction due to the relatively large size of $\Delta\Gamma_s$, of $(3.5\pm 0.2)\times 10^{-9}$ \cite{DeBrun}. The SM branching fraction for the corresponding $\overline{B}^0$ decay is even smaller due to the smallness of the CKM element $|V_{td}|$ compared with $|V_{ts}|$. The SM diagrams are shown in Fig.~\ref{Bs-mumu-diags}, along with possible interfering diagrams from NP processes. We note that many NP models are possible not just the one example of super-symmetry that is shown.
While upper limits on this process have been established by several experiments, as shown in Fig.~\ref{UL}, a new LHCb result sees evidence of a signal \cite{LHCb-Bsmumu}.\footnote{Most LHCb results reviewed here use a 1.0~fb$^{-1}$ data sample. This result uses 2.1~fb$^{-1}$.}

\begin{figure} [hbt]
\begin{center}
\vspace{-30mm}
\includegraphics[width=1.\textwidth]{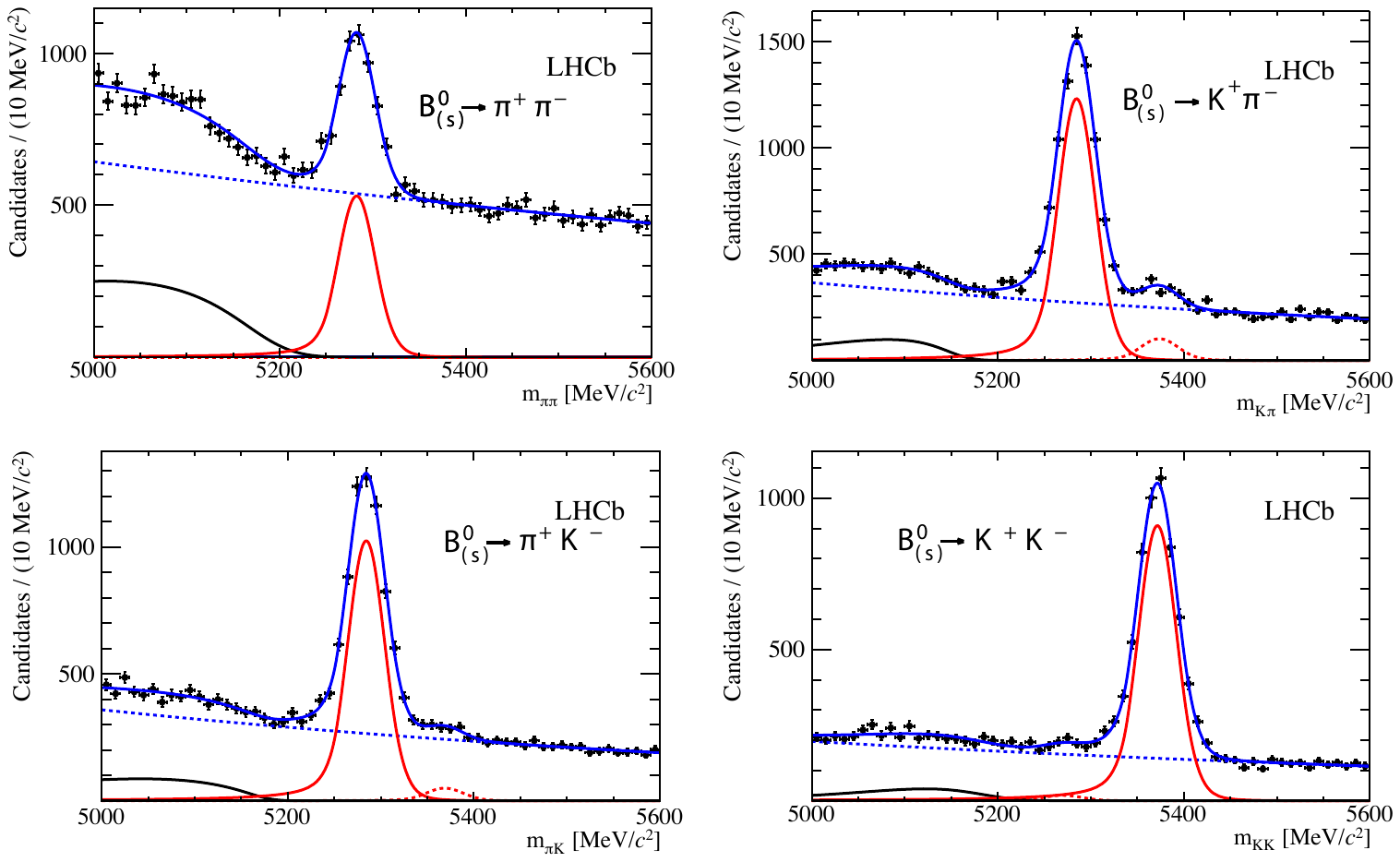} 
\vspace{-117mm}
\caption{Invariant mass spectra of the two final state hadrons for $\overline{B}_{(s)}^0\to h^+h'^-$ from different final states. Fits are shown to signal and background shapes. Signals exist for both  $\overline{B}^0$ and  $\overline{B}_s^0$ decays in some cases.} \label{LHCb-Btohh} 
\end{center}
\end{figure}

LHCb has unique capabilities at the LHC to study purely hadronic decays of $b$-flavoured hadrons. In the case of this analysis such decays are used to study selection criteria, backgrounds and to measure the transverse momentum, $p_T$, dependence of the ratio of $B_s$ to $B^0$ production, called $f_s/f_d$.
Multivariate discriminants are trained using several variables, including track impact
parameters, $B$ lifetime, $B$ $p_T$, $B$ isolation, muon isolation, minimum
impact parameter of muons, etc... The signal properties are established using fully reconstructed $B$ decays into two oppositely charged hadrons. The signals for four different final states are shown in Fig.~\ref{LHCb-Btohh}.

\begin{figure}[htb]
  \begin{center}
    \includegraphics*[width=0.45\textwidth]{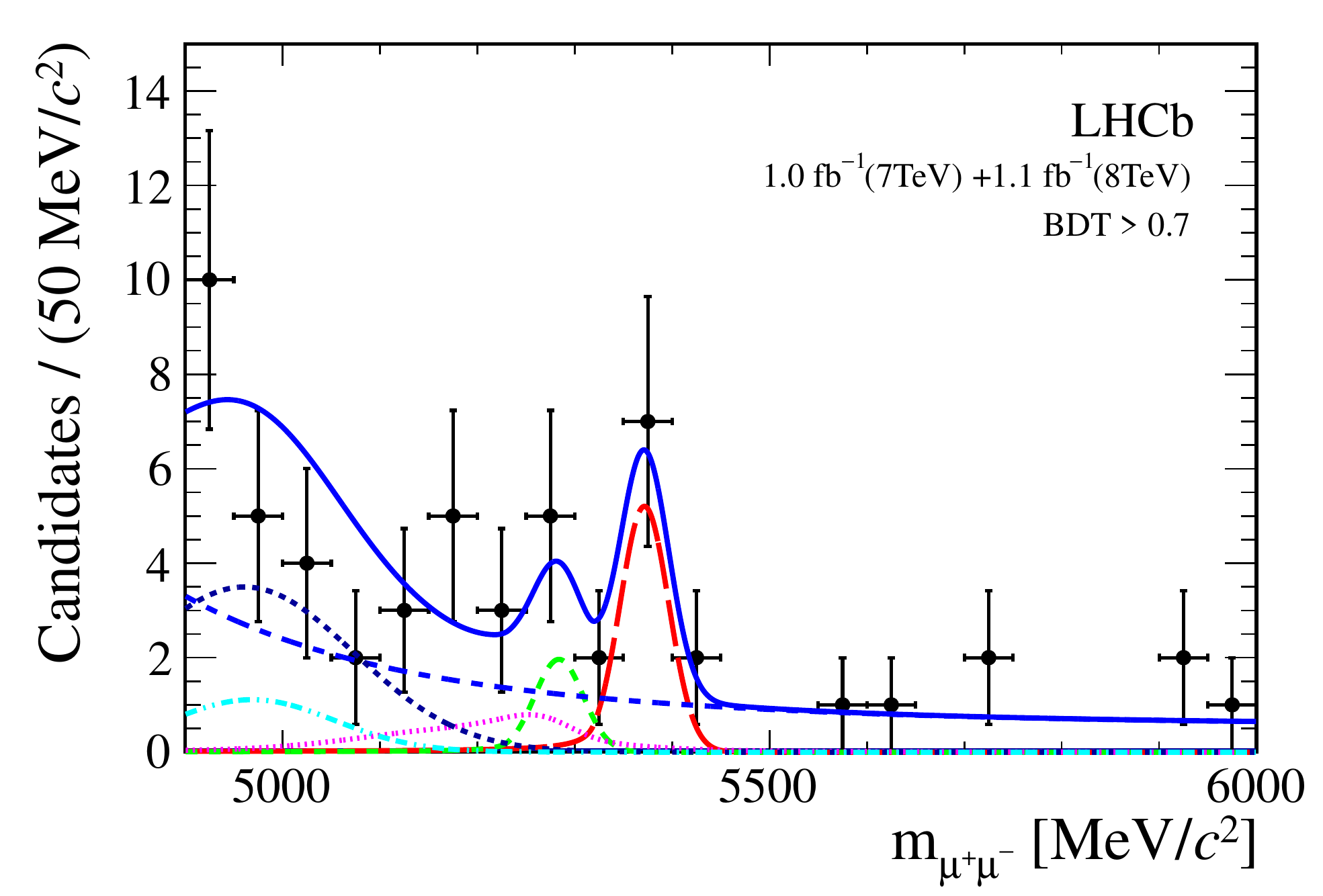}
  \end{center}
  \vspace{-5mm}
\caption{
Invariant  $\mu^+\mu^-$ mass distribution of the selected  candidates (black dots) with ${\rm BDT}>0.7$. The result of the fit is overlaid (blue solid line) and the different components detailed:
$\overline{B}_s^0$  (red long dashed), $\overline{B}^0$ (green medium dashed), $\overline{B}_s^0\to h^+h'^-$ (pink dotted), $\overline{B}^0\to\pi^-\mu^+\nu$ (black short dashed) and $\overline{B}\to\pi\mu^+\mu^-$ 
(light blue dot dashed), and the combinatorial background (blue medium dashed).}
\label{Bstomumu-sig}
\end{figure}

The analysis is done applying two different multivariate discriminates in sequence. The last one is a Boosted Decision Tree (BDT), which is constructed to have the signal probability flat in BDT, and the background peaked around zero. The data show a 3.5 standard deviation significance signal from the final fit. The projection for the final boosted decision tree variable, BDT$>$0.7 is shown in Fig.~\ref{Bstomumu-sig}.

\begin{figure}[htb]
  \begin{center}
    \includegraphics*[width=0.45\textwidth]{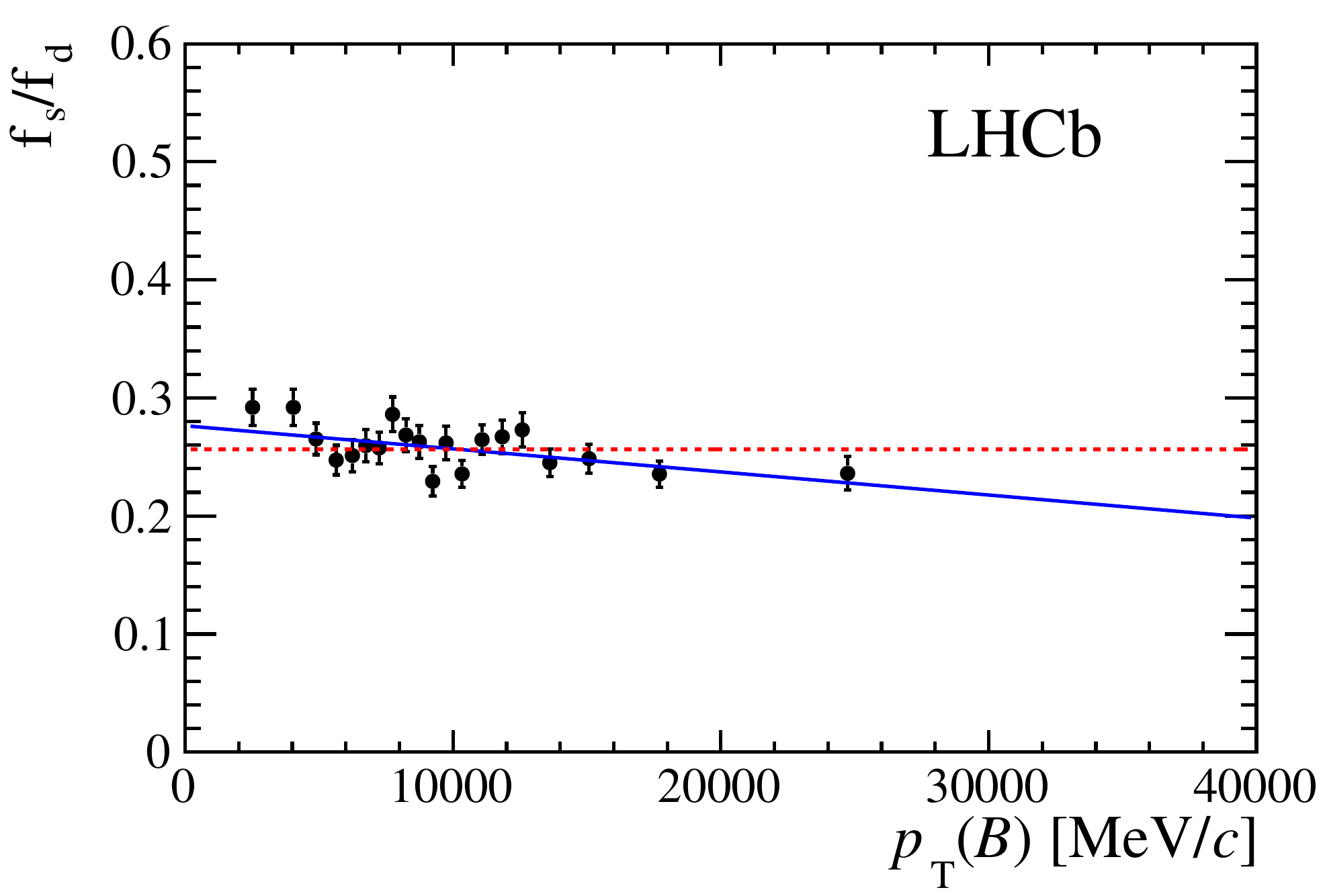}
  \end{center}
  \vspace{-5mm}
\caption{The variation of $f_s/f_d$ as a function of $p_T$ measured by LHCb using fully
reconstructed hadronic decays.}
\label{pt_dependency}
\end{figure}

To find a branching fraction one needs to determine not only the number of signal events and their average detection efficiency, but also the total number of $\overline{B}_s^0$ events. This number is inferred by measuring the number of events in  the similar decay channels  $B^-\to J/\psi K^-$ and $\overline{B}^0\to K^-\pi^+$, using their known branching fractions, and knowledge of the production ratio for $\Bsb/\overline{B}^0$ decay, $f_s/f_d$ (assuming that $B^-$ and $\overline{B}^0$ production are equal.)   LHCb determines this ratio from two sources. The most precise is semileptonic $b$ decays \cite{fsfdsemi}, and the second is specific hadronic channels using a theoretical model. The average of the two methods gives $f_s/f_d=0.256\pm 0.020$ \cite{fsfdhad}.\footnote{The hadronic channels used are $\overline{B}^0\to D^+K^-$,  $\overline{B}^0\to D^+\pi^-$ and $\Bsb\to D_s^+K^-$.} The hadronic channels also allows an exploration of the pseudo-rapidity, $\eta$, and $p_T$ dependence. It is found to be independent of $\eta$, but in $p_T$ the $\Bsb/\overline{B}^0$ yield decreases as shown in Fig.~\ref{pt_dependency}.  For LHCb the $p_T$ distribution of the signal is quite similar so little systematic uncertainty is introduced.
Other LHC experiments have in the past used the LHCb value for $f_s/f_d$; they will have to take into account this $p_T$ dependency in the future. 

\begin{figure}[htb]
  \begin{center}
    \includegraphics*[width=1.0\textwidth]{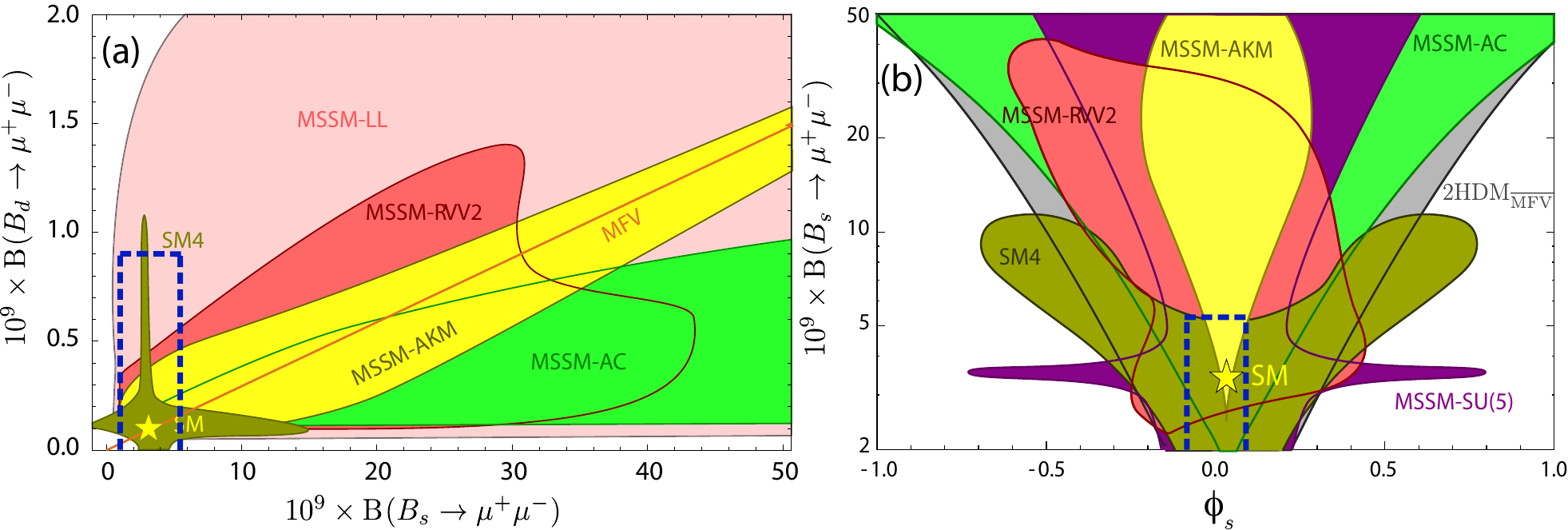}
  \end{center}
  \vspace{-3mm}
\caption{(a) Correlation between  ${\cal{B}}(\overline{B}_s^0\to\mu^+\mu^-)$ and  ${\cal{B}}(\overline{B}^0\to\mu^+\mu^-)$ in minimum flavor violation (MFV) model, the four-generation
SM4  model, and four SUSY models.  Correlation between  ${\cal{B}}(\overline{B}_s^0\to\mu^+\mu^-)$ and the mixing-induced CP
asymmetry $\phi_s$ in SM4, the two-Higgs doublet model with 
flavour blind phases, and three SUSY  models. In both cases, the SM point is marked by a star. The gray area is ruled out experimentally. In both (a) and (b) the blue dashed region outlines the allowed region within one standard deviation of their measured values. Adapted from D. Straub \cite{Straub}.
}
\label{Straub1}
\end{figure}

The most important implication of this measurement is that it is consistent with the SM prediction. It rules out many SUSY models especially those with large values of the parameter $\tan\beta$ \cite{Mahmoudi}.
Many NP models are eliminated or severally constrained when looked combined with other measurements. For example,  the concurrently available upper limit on ${\cal{B}}(\overline{B}^0\to\mu^+\mu^-) <0.94\times 10^{-9}$ at 95\% CL is plotted versus ${\cal{B}}(\overline{B}_s^0\to\mu^+\mu^-)$  for several models including the SM in Fig.~\ref{Straub1}(a). Here the space taken by any particular model is caused by a ``reasonable'' variation of its parameters. In Fig.~\ref{Straub1}(b) the models are plotted in $\phi_s$ versus ${\cal{B}}(\overline{B}_s^0\to\mu^+\mu^-)$ plane. Many models are gone or mostly eliminated. SM4, a fourth-generation model, is eliminated by the Higgs discovery. Here the 4th generation quarks would cause the Higgs production cross-section to be nine times larger and suppress the decay into $\gamma\gamma$ \cite{SM4death}. Other models with multiple Higgs doublets are allowed for heavy fourth-generation quarks with masses larger than 400 GeV \cite{Soni4th}.

\subsection{\boldmath $\overline{B}\to K^{*}\ell^+\ell^-$}

\begin{figure}[htb] 
  \begin{center}
    \includegraphics*[width=0.9\textwidth]{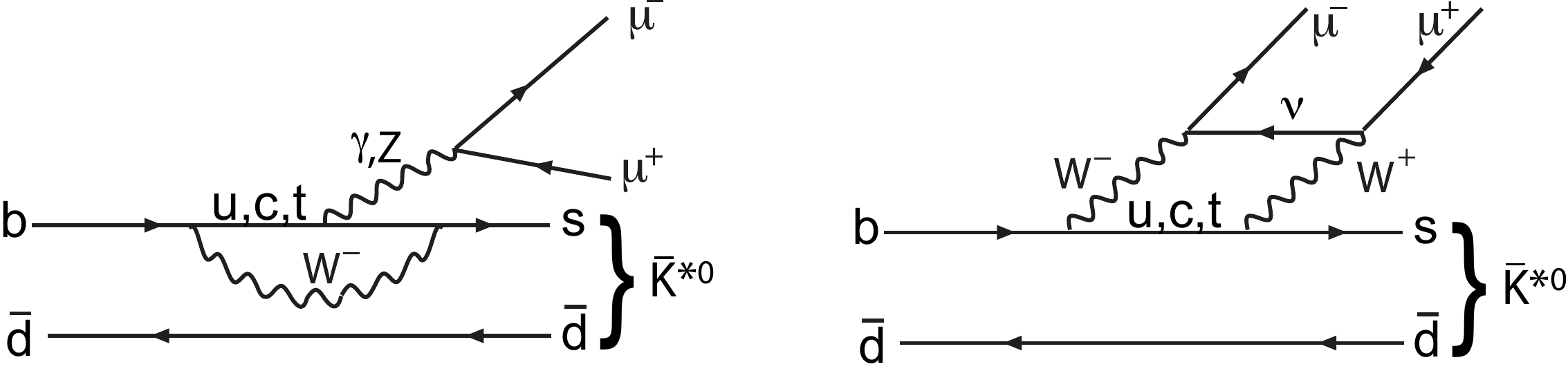}
  \end{center}
\caption{Decay diagram for $\overline{B}^0\to \overline{K}^{*0}\mu^+\mu^-$. The isospin related diagram for $B^-$ decays is almost identical, as is the one for $e^+e^-$ in the final state.
}
\label{B-K_mumu}
\end{figure}

The process $\overline{B}\to \overline{K}^{*}\ell^+\ell^-$ is well worth studying as it proceeds via two SM model loop diagrams, thus allowing for substantial NP contributions, and it is well predicted in the SM \cite{K*mumu-pred}. The diagram for the specific decay channel  $\overline{B}^0\to \overline{K}^{*0}\mu^+\mu^-$ is shown in Fig.~\ref{B-K_mumu}. Although this is an exclusive hadronic decay as contrasted with corresponding inclusive decay $b\to s \ell^+\ell^-$, there is sufficient precision in the theoretical predictions to make this an exceeding interestly mode to study \cite{K*mumu-pred}, especially useful as the exclusive decay is much more experimentally accessible. Indeed,  there are many suggestions for variables to use to uncover NP in this decay. Here I will use just $q^2$ the four-momentum transfer between the $B$ and the $K^*$, and the forward-backward asymmetry, $A_{\rm FB}$, defined as the angle of the $\mu^+$ with respect to the $B$ direction in the di-muon rest frame.
An essentially identical diagram can be drawn for the $\overline{K}\mu^+\mu^-$, but the spin-0 $K$ provides a less interesting final state in terms of decay amplitudes. 

\begin{figure}[htb] 
  \begin{center}
 \vspace{-80mm}
    \includegraphics*[width=0.8\textwidth]{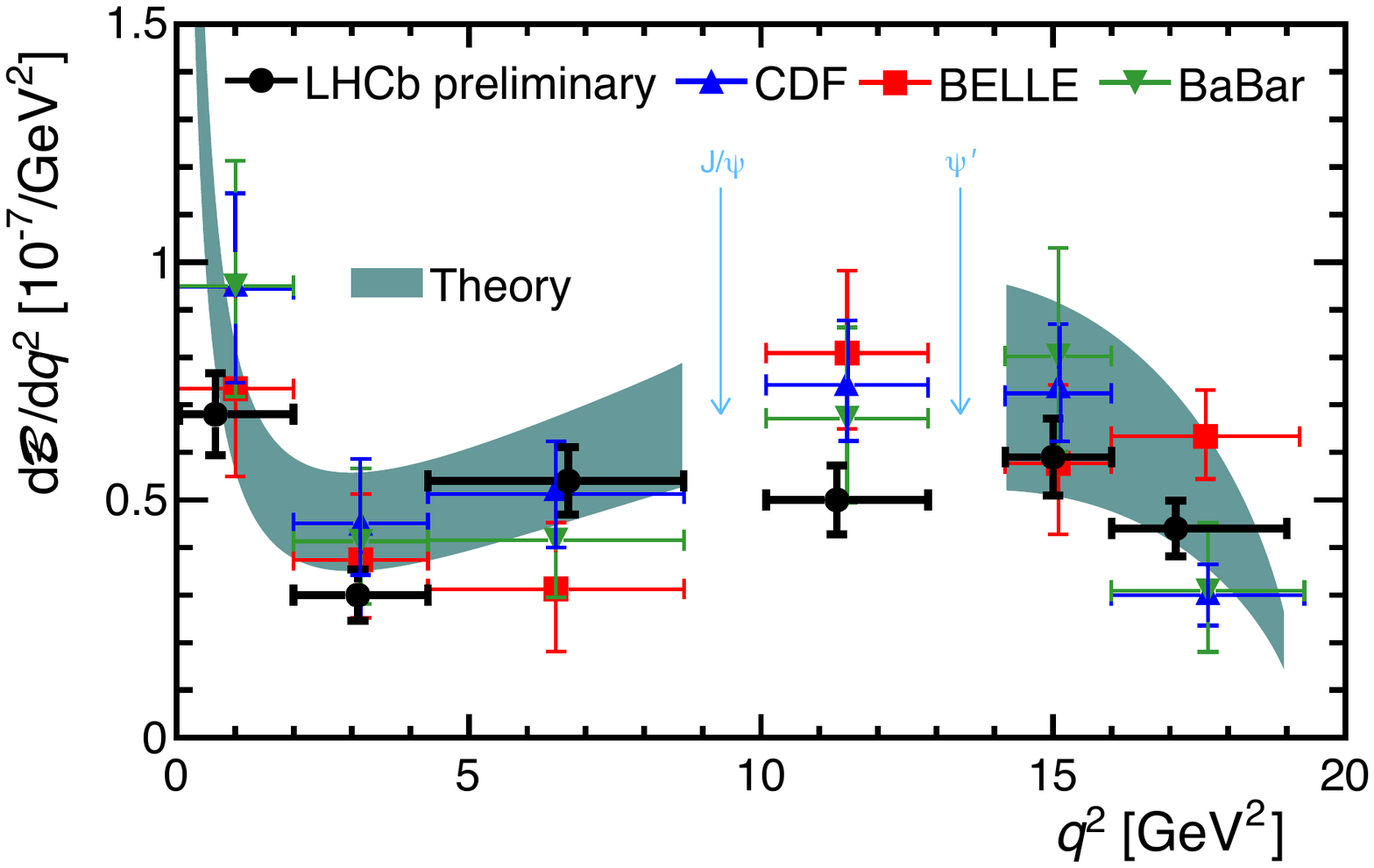}
  \end{center}
 \vspace{-25mm}
\caption{Measured decay rates for $\frac{d{\cal{B}}}{dq^2}\left(\overline{B}^0\to \overline{K}^{*0}\mu^+\mu^-\right)$ compared to a theoretical calculation from Bobeth \etal ~\cite{Bobeth-Ksmumu}.
}
\label{Kstarmumu-dGdq2}
\end{figure}

The decay rates as a function of $q^2$ from four experiments BaBar \cite{BaBar-Ksmumu}, Belle \cite{Belle-Ksmumu},
CDF \cite{CDF-Ksmumu}. and LHCb \cite{LHCb-Ksmumu} are shown in Fig.~\ref{Kstarmumu-dGdq2} and compared with a theoretical prediction \cite{Bobeth-Ksmumu}. All four experiments agree with the prediction; LHCb has the most precise measurements.

\begin{figure}[htb] 
  \begin{center}
    \includegraphics*[width=0.7\textwidth]{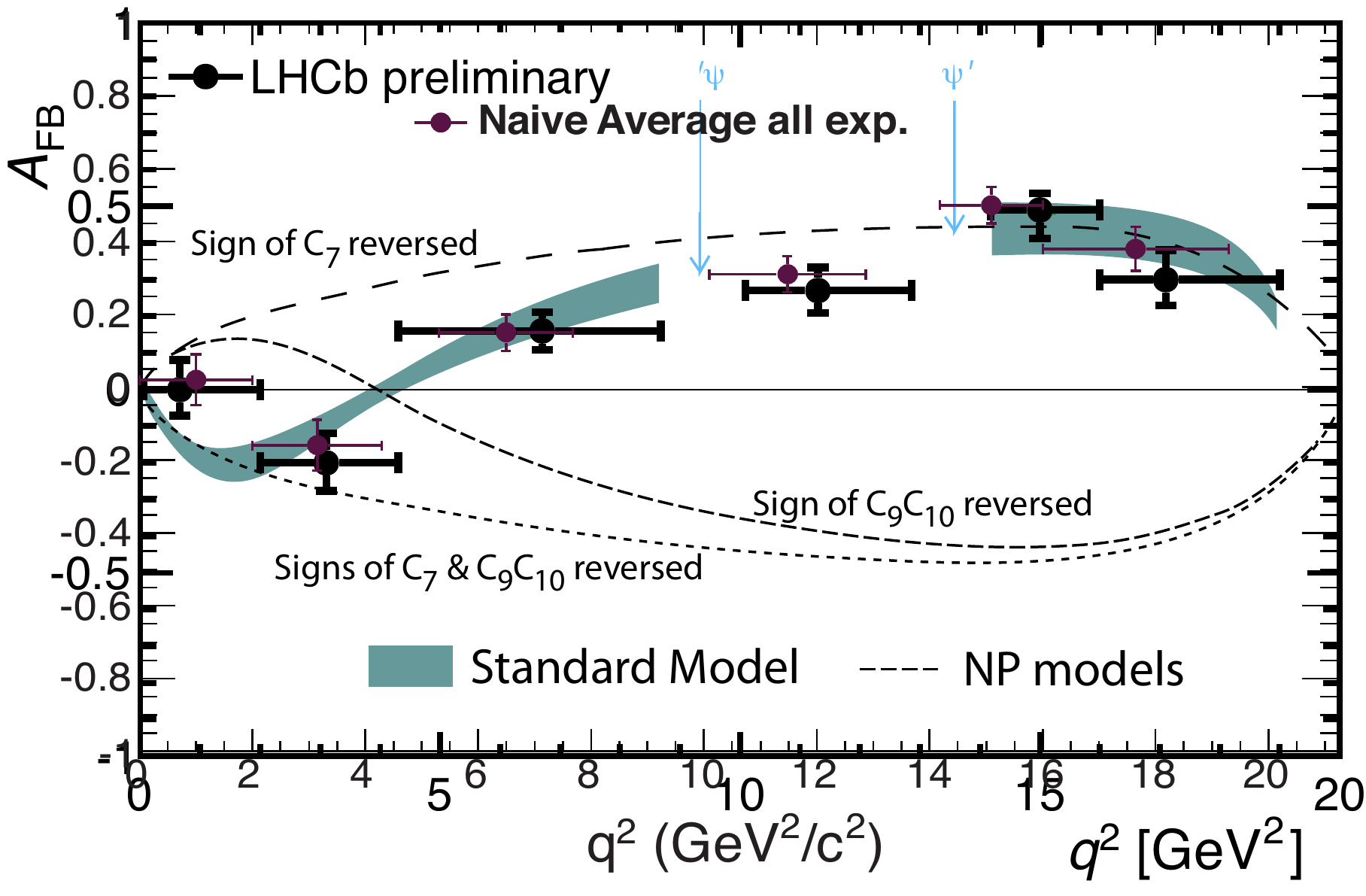}
  \end{center}
\caption{Data and theory comparison of the forward-backward asymmetry, $A_{\rm FB}$ for $\overline{B}^0\to \overline{K}^{*0}\mu^+\mu^-$ decay. 
The data are from LHCb, and the theory curves are from Bobeth \etal~\cite{Bobeth-Ksmumu}. The SM prediction is shown shaded with the band covering $\pm 1$ standard deviation uncertainty estimates. The non-SM curves are labeled with $C_i$, the meaning of which will be discussed in the next section.
Also shown is the data averaged over all four experiments, where the horizontal axis is shifted
somewhat with respect to the LHCb data. 
}
\label{Ksmumu-Afb-models-all}
\end{figure}

$A_{\rm FB}$ is more sensitive to NP. The LHCb results and an average of the data from all four experiments shown in Fig.~\ref{Ksmumu-Afb-models-all}, however, are quite consistent with the SM prediction. LHCb also measure the point at which $A_{\rm FB}$ crosses zero as $(4.9^{+1.1}_{-1.3})$~GeV$^2$, where the SM predictions range from 4.0 to 4.3 GeV$^2$ \cite{AFB-zero}.

\subsection{\boldmath Other rare $b\to s$ processes}
\label{sec:rare-s}
Other processes dominated by the $b\to s$ transition can be combined with the previously discussed ones in a common analysis framework. The formalism developed starts with writing the effective Hamiltonian as \cite{Bobeth-btosg-form}
\begin{equation}
\label{eq:Heff}
{\cal H}_{\rm eff} = - \frac{4\,G_F}{\sqrt{2}} V_{tb}V_{ts}^* \frac{e^2}{16\pi^2}
\sum_i
(C_i O_i + C'_i O'_i) + \text{h.c.}~,
\end{equation}
where the $C_i$ and $O_i$ represent SM coefficients and operators, and the $C^{\prime}_i$ and $O^{\prime}_i$ represent NP coefficients and operators. 
The operators most sensitive to NP effects are 

\begin{align}
\label{eq:O7}
O_7 &= \frac{m_b}{e}
(\bar{s} \sigma_{\mu \nu} P_R b) F^{\mu \nu},
&
O_8 &= \frac{g m_b}{e^2}
(\bar{s} \sigma_{\mu \nu} T^a P_R b) G^{\mu \nu \, a},
\nonumber\\
O_9 &= 
(\bar{s} \gamma_{\mu} P_L b)(\bar{\ell} \gamma^\mu \ell)\,,
&
O_{10} &=
(\bar{s} \gamma_{\mu} P_L b)( \bar{\ell} \gamma^\mu \gamma_5 \ell)\,,
\nonumber\\
O_S &= 
m_b (\bar{s} P_R b)(  \bar{\ell} \ell)\,,
&
O_P &=
m_b (\bar{s} P_R b)(  \bar{\ell} \gamma_5 \ell)\,,
\end{align}
where $m_b$ denotes the running $b$ quark mass in the $\overline{\rm MS}$
scheme, $P_L=(1-\gamma_5)/2$, and $P_R=(1+\gamma_5)/2$. The NP operators $O^{\prime}$ are found by switching $P_R$ for $P_L$ and vice-versa  \cite{Alt-Straub}.

There are other rare $b\to s$ processes that contribute to the NP search. One important one is the exclusive decay $\overline{B}^0\to K^{*0}\gamma$, $K^{*0}\to K_s\pi^0$. Time dependent CP violation for this process is given by
\begin{equation}
\frac{\Gamma\left(\overline{B}^0(t)\to K^{*0}\gamma\right)-\Gamma\left({B}^0(t)\to K^{*0}\gamma\right)}{\Gamma\left(\overline{B}^0(t)\to K^{*0}\gamma\right)+\Gamma\left({B}^0(t)\to K^{*0}\gamma\right)}=S_{K^*\gamma}\sin\left(\Delta m_d t\right) - C_{K^*\gamma}\cos\left(\Delta m_d t\right)~.
\end{equation}

Including generic NP operators, the expected asymmetry is parameterized as
\begin{equation}
S_{K^*\gamma}=\frac{2}{\left|C_7\right|^2+\left|C^{\prime}_7\right|^2}{\rm Im}\left(e^{-2i\beta}C_7C^{\prime}_7\right).
\end{equation}

The SM prediction for $S_{K^*\gamma}$, where $C^{\prime}_7=0$, is $(-0.023\pm 0.016)$, while the measurement from BaBar and Belle \cite{HFAG-btosg} is $-0.16\pm0.22$. Clearly reducing the error would be useful, but the measurement, as we will see, is already useful.

Another useful process is the inclusive decay $b\to s \ell^+\ell^-$. Here two four-momentum transfer, $q^2$, regions are defined, one at low values $1<q^2<6$~GeV$^2$ which again probes $C^{\prime}_7$ and the other with $q^2>14.4$~GeV$^2$ which examines both $C^{\prime}_9$ and $C^{\prime}_{10}$. This is similar to the coefficients probed in the exclusive $\overline{B}^0\to \overline{K}^{*0}\ell^+\ell^-$ channel. 

\begin{figure}[htb] 
  \begin{center}
    \includegraphics*[width=0.9\textwidth]{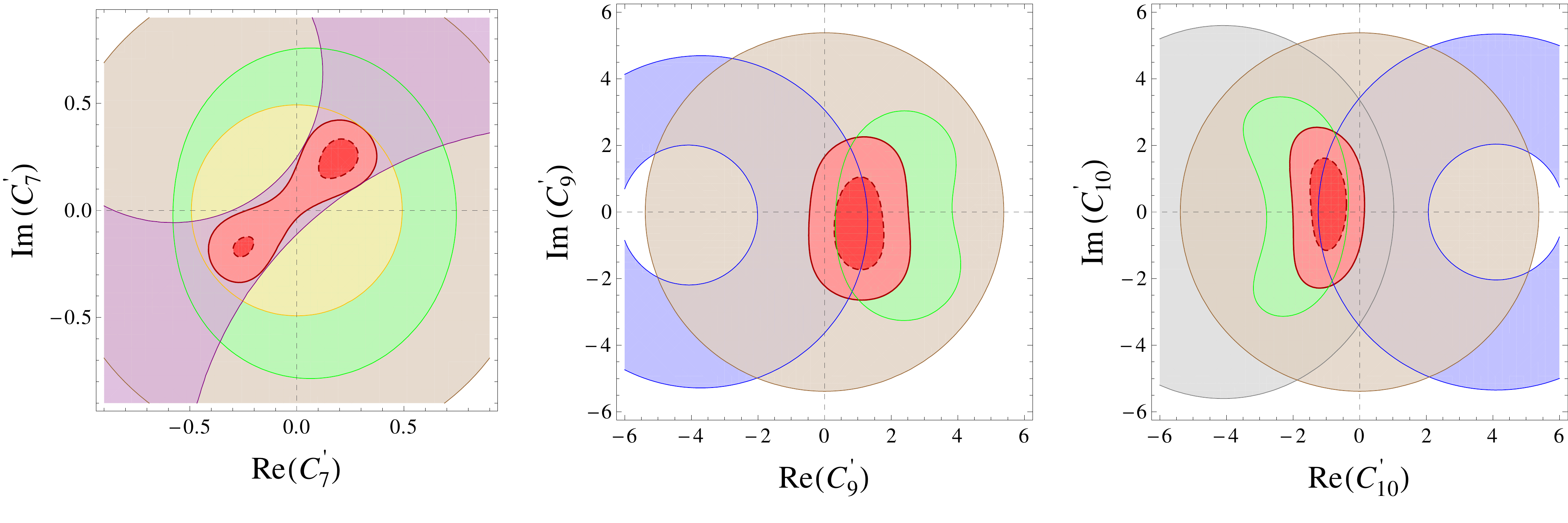}
  \end{center}
 \vspace{-8mm}
\caption{Individual two standard deviation constraints on the primed Wilson coefficients as well as combined one and two standard deviation constraints  shown in red (dark and light) from $B\to X_s\ell^+\ell^-$ (brown), ${\cal{B}}(B\to X_s\gamma$) (yellow), $A_\text{CP}(b\to s\gamma)$ (orange), $B\to K^*\gamma$ (purple), $B\to K^*\mu^+\mu^-$ (green), $B\to K\mu^+\mu^-$ (blue) and $\overline{B}_s\to\mu^+\mu^-$ (gray) from \cite{Alt-Straub} .
}
\label{bandplotsRH}
\end{figure}

Putting all of the channels together Altmannshofer and Straub provide constraints on the real and imaginary parts of the NP parameters shown in Fig.~\ref{bandplotsRH}. The SM points are at (0,0), consistent with the analysis within $2\sigma$ in all cases.

\section{Tree level decays with ``abnormal'' behavior}
\subsection{\boldmath $B^-\to \tau^- \overline{\nu}$}

\begin{figure}[htb] 
  \begin{center}
    \includegraphics*[width=0.4\textwidth]{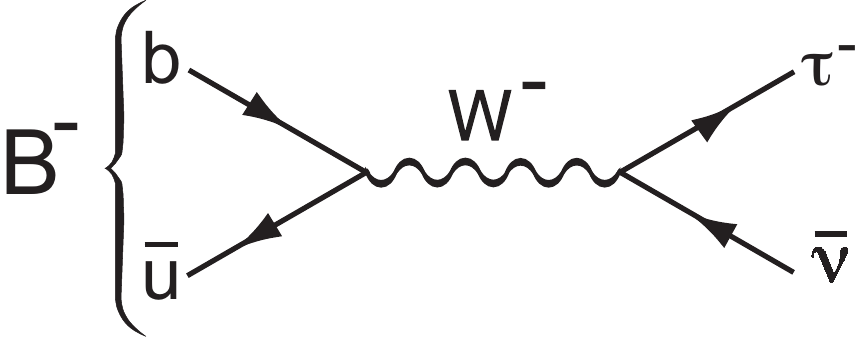}
  \end{center}
 \vspace{-6mm}
\caption{Feynman diagram for ${\cal{B}}(B^-\to \tau^-\overline{\nu})$.
}
\label{Btotaunu}
\end{figure}

Here we depart from the paradigm of considering NP via loop diagrams and look instead at a the suppressed diagram shown in Fig.~\ref{Btotaunu}. The branching fraction in the SM is given by
\begin{equation}
{\cal{B}}(B^-\to \tau^-\overline{\nu})=\frac{G_F^2m_{B^-}\tau_{B^-}}{8\pi}m_{\tau}^2\left(1-\frac{m_{\tau}^2}{m_{B^-}^2}\right)^2f_B^2\left|V_{ub}\right|^2,
\end{equation}
where $V_{ub}$ is a small CKM element and $f_B$ is the $B$ meson decay constant. Helicity suppression, caused by having the spin-0 $B^-$ decay into a left-handed lepton and a right-handed anti-lepton whose helicities align for massless leptons and thus tend to form a spin-1 system, is partially broken due to the relative large value of $m_{\tau}$ with respect to $m_{B^-}$. New physics could enter here by having another charged boson, e.g. a virtual $H^-$ also participate in the decay instead of the $W^-$ \cite{Btotaunu-NP}.

The measurements of the branching fraction have been made by BaBar and Belle. Both collaborations have separate determinations where they either fully reconstruct a hadronic decay or a semi-leptonic decay tag, and look for evidence for a $\tau^-$ decay on the other side requiring little extra energy in the event after accounting for the tag and the $\tau^-$ decay products.
The experimental measurements have been compiled by Rosner and Stone \cite{Rosner-Stone} and listed in Table~\ref{tab:Btotaunu}.
\begin{table}[htb]
\label{tab:Btotaunu}
\begin{center}
\begin{tabular}{lllc} \hline\hline
&Experiment & Tag &${\cal{B}}$ (units of $10^{-4}$)\hfill\\
\hline
&Belle~\cite{BelleH}&Hadronic&$1.79\,^{+0.56\,+0.46}_{-0.49\,-0.51}$\\
&Belle~\cite{BelleS}&Semileptonic&$1.54\,^{+0.38\,+0.29}_{-0.37\,-0.31}$\\
&Belle&Our average&$1.62 \pm 0.40$ \\
&BaBar~\cite{BaBarH} & Hadronic & $1.80\,^{+0.57}_{-0.54}\pm0.26$\\
&BaBar~\cite{BaBarS} & Semileptonic & $1.7\pm 0.8\pm 0.2$\\
&BaBar & Average \cite{BaBarH} & $1.76 \pm 0.49$\\\hline
& &Our average & $1.68\pm0.31$\\
\hline\hline
\end{tabular}
\caption{Experimental results for ${\cal{B}}(B^-\to \tau^-\overline{\nu})$ prior to this conference from Rosner and Stone \cite{Rosner-Stone}.}
\end{center}
\end{table}
The Belle average was constructed assuming that the systematic uncertainties in the two measurements are uncorrelated, while BaBar supplies their average. The measurements are all very consistent, perhaps too consistent, considering the quoted uncertainties. 

\begin{figure}[htb] 
  \begin{center}
    \includegraphics*[width=0.6\textwidth]{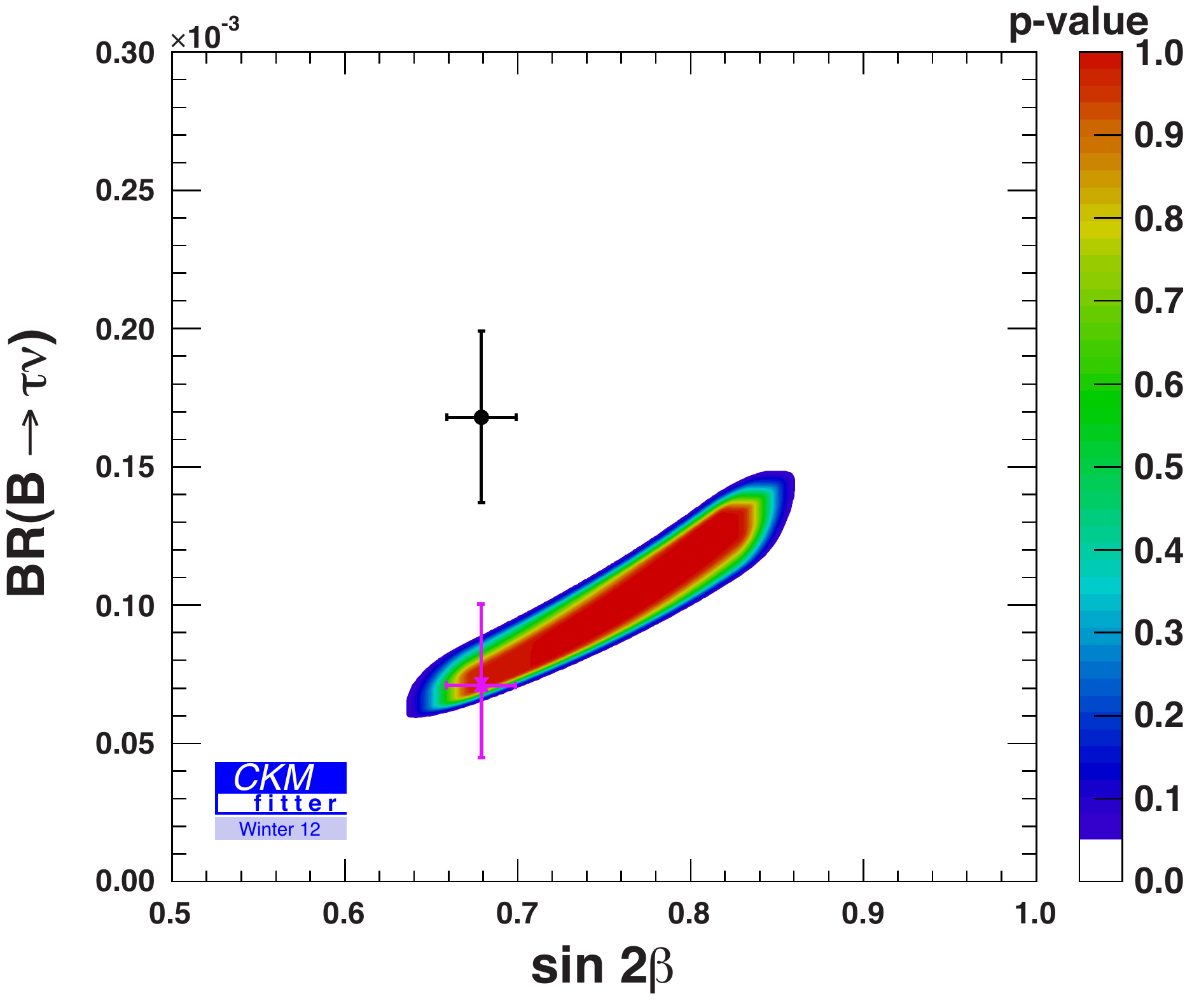}
  \end{center}
 \vspace{-8mm}
\caption{Allowed correlated values of ${\cal{B}}\left(B^-\to \tau^-\overline{\nu}\right)$ and $\sin 2\beta$, shown in the banana like shape,  compared with the experimental averages, shown as the small black circle, prior to this conference. The pink x shows the newly measured value from Belle for hadronic tags only. In all cases the errors are a combination of statistical and systematic uncertainties.
}
\label{tension_sin2b_Btaunu}
\end{figure}

Prior to this conference one inconsistency stood out that could be the first sign of NP.  The CKM fitter \cite{CKMfitter} and UTfit \cite{UTfit} groups had explored the consistency of all processes, by relating the the CKM constants $A$, $\lambda$, $\rho$ and $\eta$ to individual measurements, and deriving values for these numbers. While the overall consistency was at the $\sim\!2\sigma$ level,  delving into the individual measurements showed a looming difference between the value of the $CP$ violating asymmetry in $\overline{B}^0\to J/\psi K_S^0$ decays, $\sin 2\beta$, and the value of ${\cal{B}}\left(B^-\to \tau^-\overline{\nu}\right)$. The CKM fitter group showed this discrepancy nicely by predicting the values of $\sin 2\beta$ and ${\cal{B}}\left(B^-\to \tau^-\overline{\nu}\right)$ without using either measurement and comparing the prediction with actual measurements; see Fig.~\ref{tension_sin2b_Btaunu}. 
The experimental average for $\sin 2\beta$ is consistent with the predicted value, while the pre-conference value of ${\cal{B}}\left(B^-\to \tau^-\overline{\nu}\right)$ is higher by a bit more than $2\sigma$. This problem was taken quite seriously by some: one paper appeared entitled ``Demise of CKM and its aftermath'' \cite{Lunghi-Soni}. As we shall see shortly, this conclusion may have been a bit premature.

\begin{figure}[htb] 
  \begin{center}
    \includegraphics*[width=0.5\textwidth]{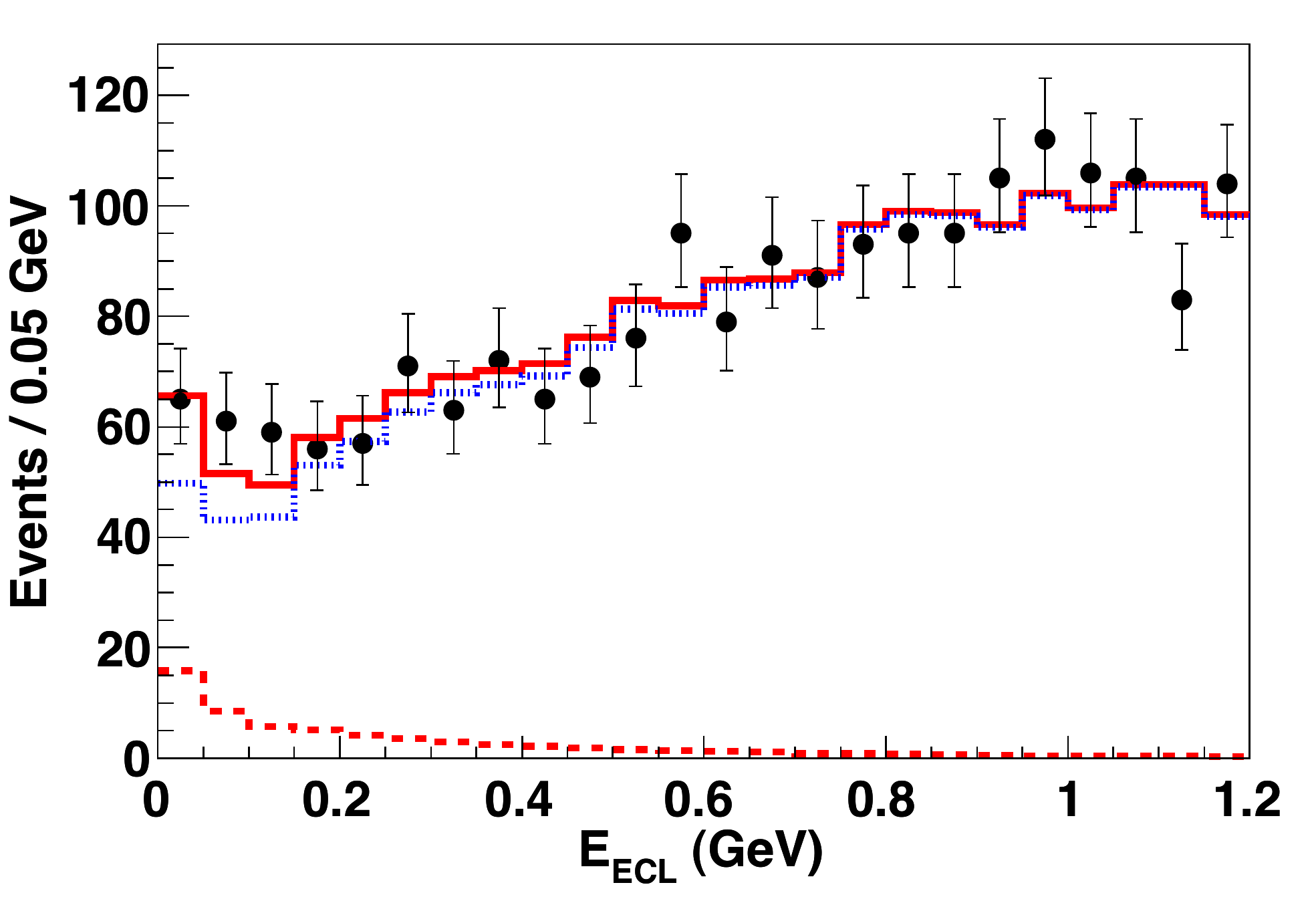}\includegraphics*[width=0.5\textwidth]{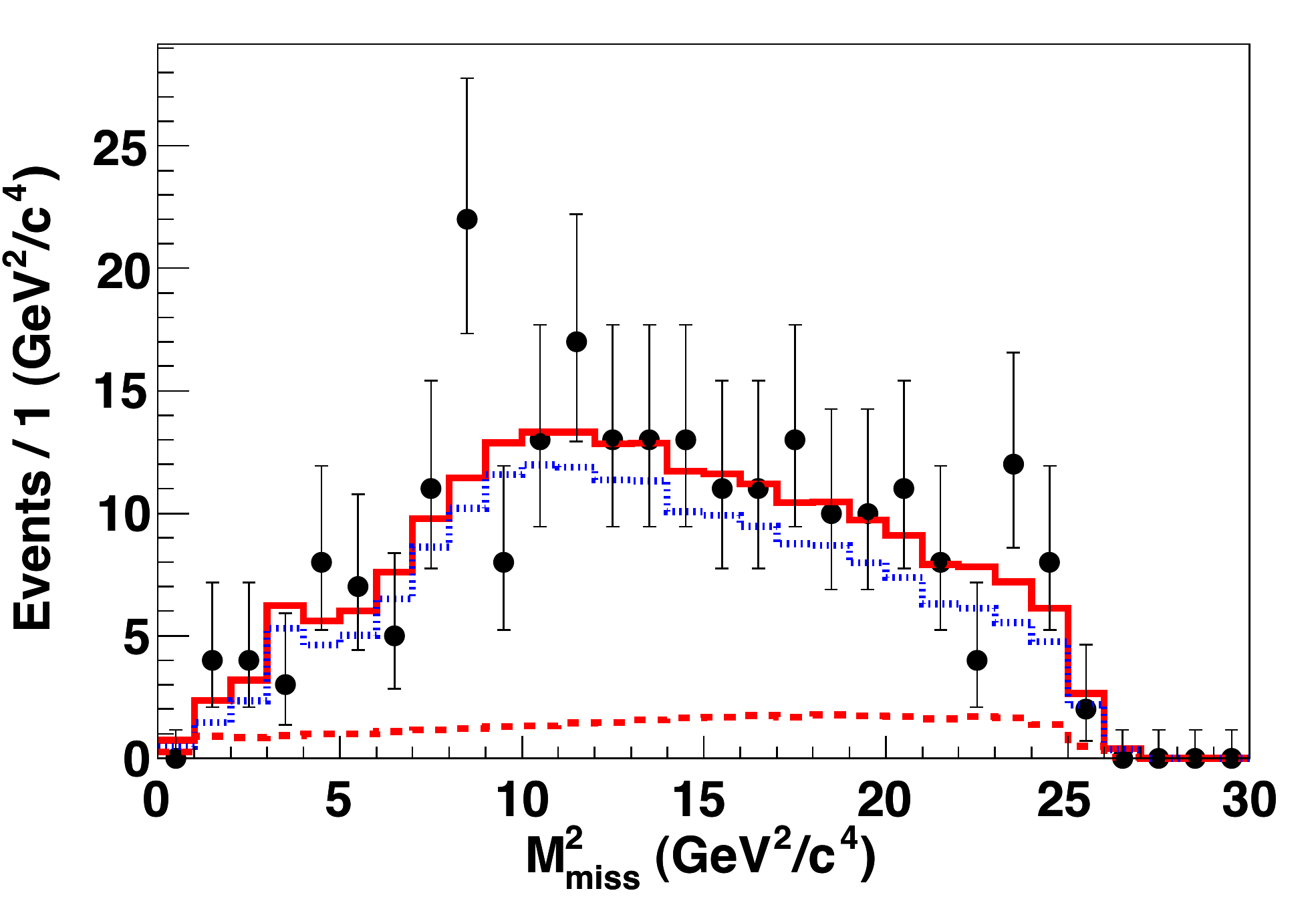}
  \end{center}
 \vspace{-4mm}
\caption{Distribution of ${\rm E}_{\rm ECL}$ (left) combined for all the $\tau^-$ decays from Belle.  The M$^2_{\rm miss}$, distribution (right) is shown for  ${\rm E}_{\rm ECL} < 0.2~{\rm GeV}$.
  The solid circles with error bars are data, while the
  the solid histograms show the projections of the fits, with
  the dashed and dotted histograms indicating the signal and background components, respectively.
}
\label{Belle-extra-E}
\end{figure}

At this conference Belle presented a new measurement using hadronic tags \cite{Belle-new-taunu}. In doing so, they made several significant advances. First of all they improved their low momentum tracking efficiency thus obtaining a factor of 2.2 improvement in the number of hadronic tags. Secondly, they increased the data sample by a factor of 1.7, they also changed their analysis technique obtaining an improved signal efficiency by a factor of 1.8, and using besides extra energy,  ${\rm E}_{\rm ECL}$, they include information from the missing mass squared, M$^2_{\rm miss}$, calculated from the hadronic tag and the $\tau^-$ decay products. These distributions are shown in Fig.~\ref{Belle-extra-E}.

The ${\rm E}_{\rm ECL}$ signal distribution peaks at zero but is widened by detector and initial state radiation effects. The M$^2_{\rm miss}$ signal distribution is easy to predict from well understood $\tau^-$ decays and is flatter than the background distribution adding to the discrimination power of the analysis. With these improvements the result is only 40\% of its previous value:  
\begin{equation}
{\cal{B}}\left(B^-\to \tau^-\overline{\nu}\right)=\left(0.72^{+0.27}_{-0.25}\pm 0.11\right)\times 10^{-4}~,
\end{equation}
being only about $3\sigma$ significant.
This branching fraction is also shown in Fig.~\ref{tension_sin2b_Btaunu} and is completely consistent with the $\sin2\beta$ measurement. We await updates using semileptonic tags from Belle as well as BaBar updates.

\subsection{\boldmath $\overline{B}\to D^{(*)}\tau^-\overline{\nu}$}
The semileptonic decay $\overline{B}\to D^{(*)}\tau^-\overline{\nu}$ proceeds in the SM via the tree-level diagram shown in Fig.~\ref{tree-loop-diag}(a), where the $c$ quark and light spectator quark form a $D$ or $D^*$ meson. As in the previous case a new charged boson could also modify this diagram especially if its coupling to the $\tau^-$ were enhanced \cite{Dtaunu-pred,Dtaunu-pred2}. 

\begin{figure}[htb] 
  \begin{center}
    \includegraphics*[width=0.8\textwidth]{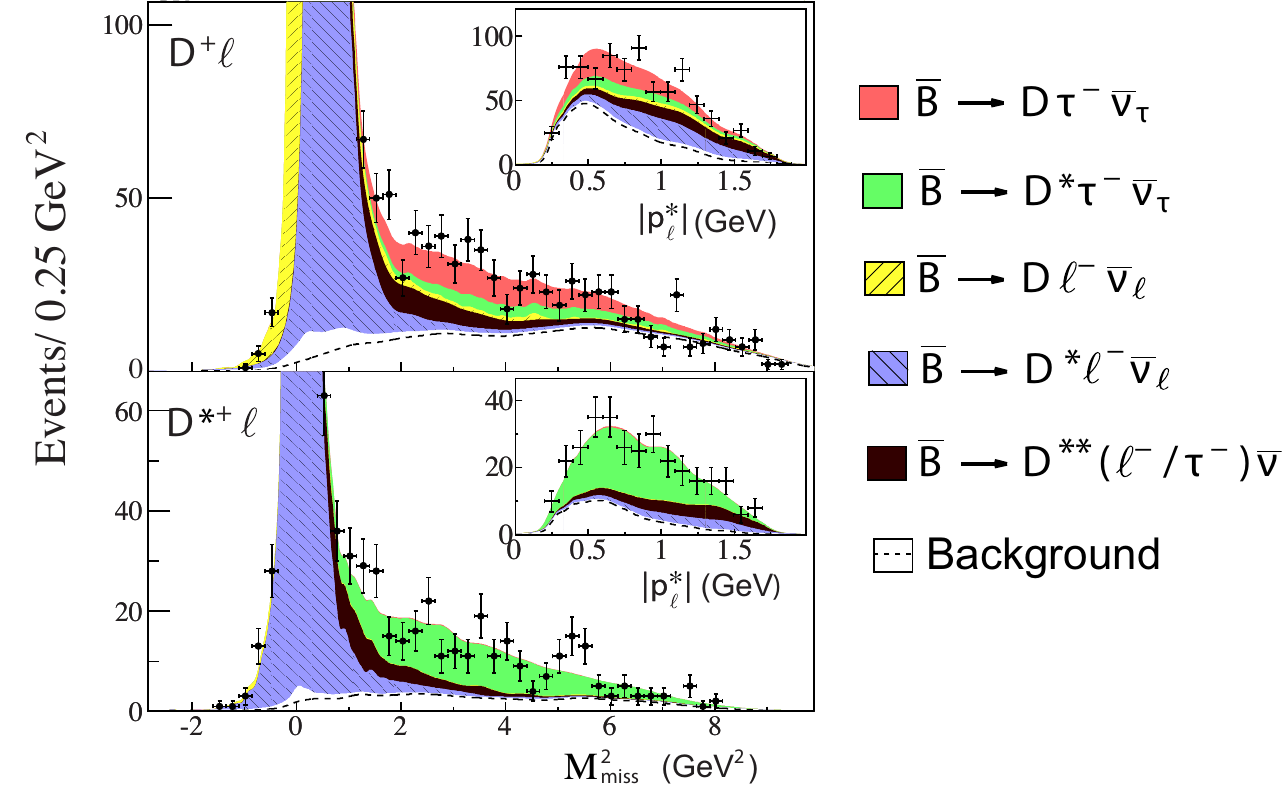}
      \end{center}
 \vspace{-4mm}
\caption{BaBar data for M$^2_{\rm miss}$ and $|p^*_{\ell}|$ (see insets) for M$^2_{\rm miss}>1$~GeV$^2$ for the $D^+\ell^-$ and $D^{*+}\ell^-$ channels.  (Data exist but are not shown for the 
$D^0\ell^-$ and $D^{*0}\ell^-$ channels.) In the background component the region above the dashed line is from charge cross-feed, while the region below contains continuum and $B\overline{B}$.
}
\label{Babar-Dtaunu}
\end{figure}

BaBar has recently shown a new analysis of these decays \cite{Babar-Dtaunu}. This analysis also requires tagging, but here only fully hadronic decay tags are used. The $\tau^-$ is observed only in its fully leptonic decays to $\mu^-$ or $e^-$, so three neutrinos are missing. The main variables used to discriminate signal from background are the M$^2_{\rm miss}$ calculated here using the tag plus the lepton, and the magnitude of the momentum of the lepton in the $B$ rest frame, $|p^*_{\ell}|$. The data are shown in Fig.~\ref{Babar-Dtaunu}. The large shaded (blue) peaks  correspond to either the $\overline{B}\to D\ell^-\overline{\nu}$ or the $\overline{B}\to D^{*}\ell^-\overline{\nu}$ channel and thus in this case constitute a background. The solid black regions correspond to $D^{**}$ production and also are a background that has a similar shape to the signal. Only the non-shaded (red) region in the top plot and the non-shaded (green) region in the bottom plot constitute signal. While it is difficult to separate the $D^{**}$ from the signal components, BaBar has used events with an extra $\pi^0$ to estimate this contribution directly from the data.

Results are reported in terms of ratios ${\cal R}(D^{(*)}) = {\cal B}(\overline{B} \rightarrow D^{(*)}\tau^-\overline{\nu})/{\cal B}(\overline{B} \rightarrow D^{(*)}\ell^-\overline{\nu})$.  These quantities are well predicted in the SM \cite{Dtaunu-pred2}. The comparison is shown in Table~\ref{tab:Babar-Dtaunu}.
\begin{table}[hbt]
\label{tab:Babar-Dtaunu}
\begin{center} 
\begin{tabular}{lccc} \hline\hline
 &SM theory & BaBar value & Difference \\\hline
${\cal R}(D)$  & $0.297\pm 0.017$ & $0.440\pm 0.058\pm 0.042$ & +2.0$\sigma$\\
${\cal R}(D^{*})$ & $0.252\pm 0.003$ & $0.332\pm 0.024\pm 0.018$&+2.7$\sigma$\\
\hline\hline
\end{tabular}
\caption{BaBar results for the ratios ${\cal R}(D^{(*)})$ compared to the SM prediction. }
\end{center}
\end{table}
Taken together, the results disagree with these expectations at 
the $3.4\sigma$ level.\footnote{The two results are inconsistent the prediction of the Type II two Higgs doublet model, because they imply different values for the model parameters.}

Belle had previously measured these ratios \cite{Belle-Dtaunu} and found excesses beyond the SM predictions; their results are listed in Table~\ref{tab:Belle-Dtaunu}. 
\begin{table}
\label{tab:Belle-Dtaunu}
\begin{center} 
\begin{tabular}{lccc} \hline\hline
 &SM theory & Belle value & Difference \\\hline
${\cal R}(D^0)$  & $0.297\pm 0.017$ & $0.70^{+0.19+0.11}_{-0.18-0.09}$ & +2.0$\sigma$\\
${\cal R}(D^+)$  & $0.297\pm 0.017$ & $0.48^{+0.22+0.06}_{-0.19-0.05}$ & +0.9$\sigma$\\
${\cal R}(D^{*0})$ & $0.252\pm 0.003$ & $0.47^{+0.11+0.06}_{-0.10-0.07}$&+1.8$\sigma$\\
${\cal R}(D^{*+})$ & $0.252\pm 0.003$ & $0.48^{+0.14+0.06}_{-0.12-0.04}$&+1.8$\sigma$\\
\hline\hline
\end{tabular}
\caption{Belle results for the ratios ${\cal R}(D^{(*)})$ compared to the SM prediction. }
\end{center}
\end{table}
Note however that the Belle results are not published, even though they were disseminated in 2009, and have rather large uncertainties compared to BaBar. It would be really interesting to have an updated analysis using the new tagging techniques that Belle uses in their most recent $B^-\to \tau^-\overline{\nu}$ paper to see if NP in these channels can be established.

\section{Other searches: the dark sector and Majorana neutrinos}
\subsection{The dark sector}

Could it be that there are 3 classes of matter? One being well known 
SM particles with charges [SU(3)xSU(2)xU(1)], the second being 
dark matter particles with ``dark'' charges, and the third being a form of matter
that has both charges, called ``mediators.''

In one such proposal  \cite{mediators} the mediator is a particle of the U(1) gauge group and is nicknamed a ``dark photon,'' playing on the oxymoronic nature of the name. In one experimental study BaBar looked for such mediators coupling to $b$ quarks \cite{Babar-dark-1S}.  The idea was to  see if there were any events where the $\Upsilon(1S)$ decayed into a real photon plus no visible energy. No such events were found and limits on branching fraction between $\approx 10^{-6}-10^{-4}$ are set, depending on the kinematical distributions of the photon.

\begin{figure}
\begin{center}
\vspace{-1mm}
\includegraphics[width=0.6\textwidth]{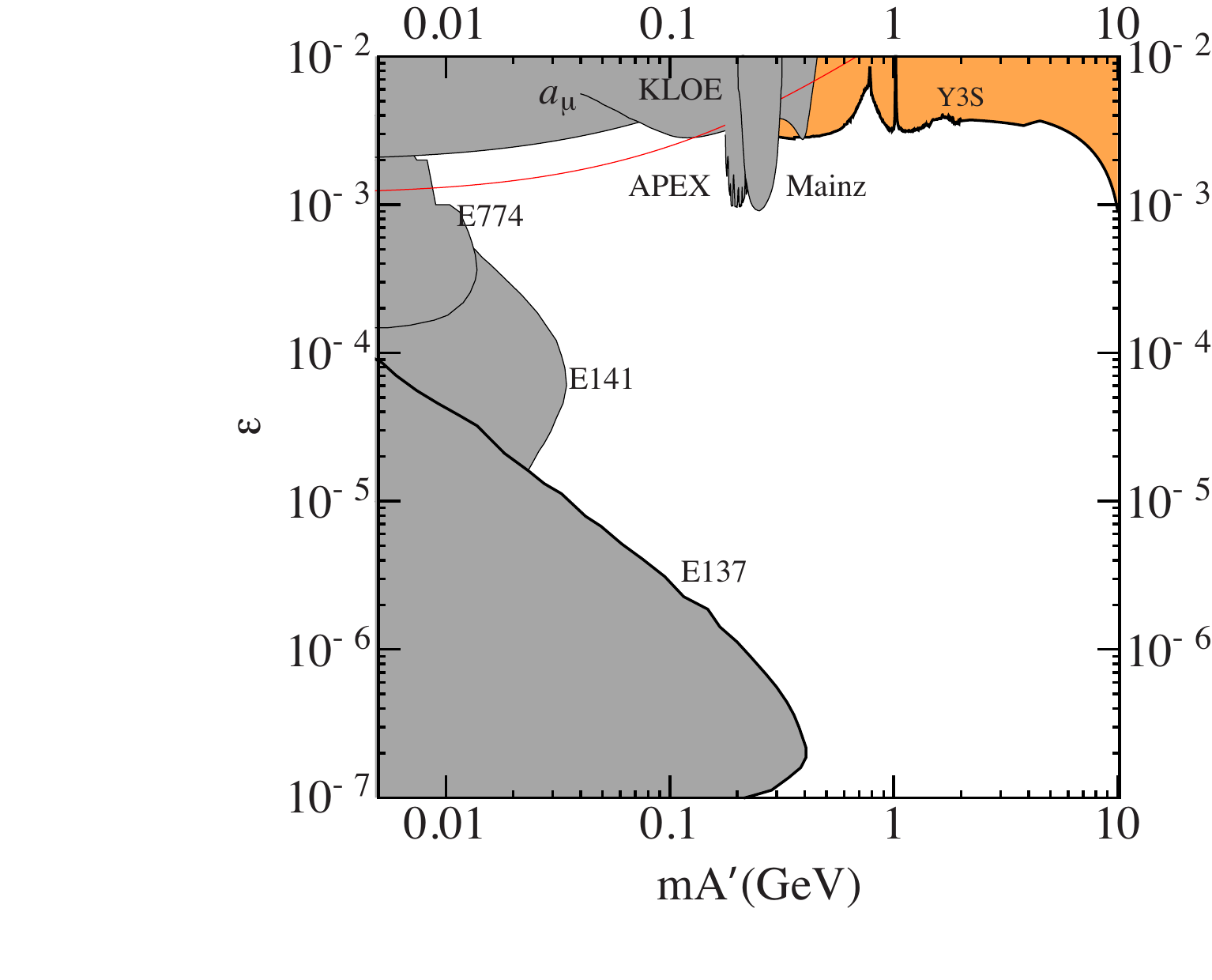}
\vspace{-5mm}
\caption{Constraints on the mixing parameters, $\epsilon$, as a function of the 
hidden photon mass derived from searches in $\Upsilon(2S,3S)$ decays at BaBar
(orange shading) and from other experiments 
\cite{mediators,Giovannella:2011nh,Abrahamyan:2011gv} (gray shading). 
The red line shows the value of the coupling required to 
explain the discrepancy between the calculated and measured anomalous magnetic 
moment of the muon~\cite{Pospelov:2008zw}. From Echenard \cite{Echenard}. }
\label{dark-limits}
\end{center}
\end{figure}

The coupling between the dark sector and the quark sector is specified by $\epsilon$, and the dark photon mass is labeled as mA$'$. Several other searches have been made and  reviewed by Echenard \cite{Echenard}. They are summarized in Fig.~\ref{dark-limits}.
There will be more experimental tests of these ideas, although these searches have already produced significant limits.

\subsection{Majorana neutrino production in $B^-$ decays}

\begin{figure}[htb] 
  \begin{center}
    \includegraphics*[width=0.8\textwidth]{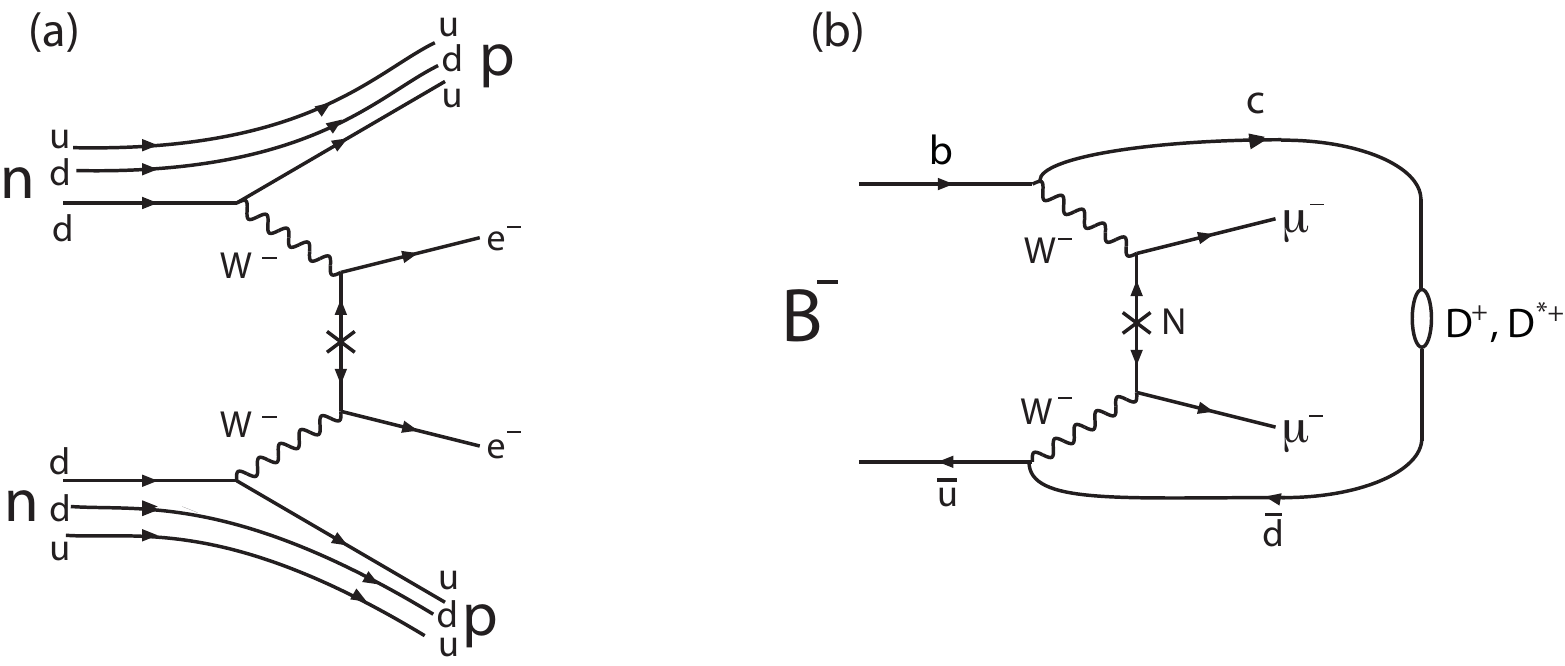}
      \end{center}
 \vspace{-.4mm}
\caption{(a) Feynman diagram for nuclear double-$\beta$ decay. (b) A related diagram for $B^-\to D^{+(*) }\mu^-\mu^-$ decays.}
\label{doublebeta-ichep}
\end{figure}

A crucial question in formulating theories beyond the SM is whether or not neutrinos are normal Dirac spin 1/2 fermions, or if they are their own anti-particles as suggested by Majorana \cite{Majorana}.
If neutrinos are indeed Majorana then they allow a process called neutrinoless double-$\beta$ decay, shown in Fig.~\ref{doublebeta-ichep}(a) \cite{doublebeta}. This process can proceed for any value of the Majorana neutrino mass as the particle is virtual in this diagram. A similar process for $B^-$ decays is shown in Fig.~\ref{doublebeta-ichep}(b). Each $\mu^-$ here can be replaced by an $e^-$ or a $\tau^-$. 
Belle \cite{Belle-Majorana} and LHCb \cite{LHCb-Majorana} have searched for these decays without success. Upper limits are given in Table~\ref{tab:Majorana1}. The upper limit in the $e^-e^-$ mode is not competitive with nuclear double-$\beta$ decay. The other modes, however, are unique since the measure coupling of the Majorana neutrino to muons.

\begin{table}
\label{tab:Majorana1}
\begin{center} 
\begin{tabular}{llc} \hline\hline
Mode & Experiment & Upper limit $\times 10^{-6}$ \\\hline
$B^-\to D^+e^-e^-$ & Belle & $<2.6$ \\
$B^-\to D^+e^-\mu^-$ & Belle & $<1.8 $\\
$B^-\to D^+\mu^-\mu^-$ & Belle & $<1.0$ \\
$B^-\to D^+\mu^-\mu^-$ & LHCb& $<0.69$ \\
$B^-\to D^{*+}\mu^-\mu^-$ & LHCb & $<3.6$ \\
\hline\hline
\end{tabular}
\caption{Upper limits at 90\% CL on $B^-$ decays to like-sign dilepton final states.}
\end{center}
\end{table}

\begin{figure}[htb] 
  \begin{center}
    \includegraphics*[width=0.7\textwidth]{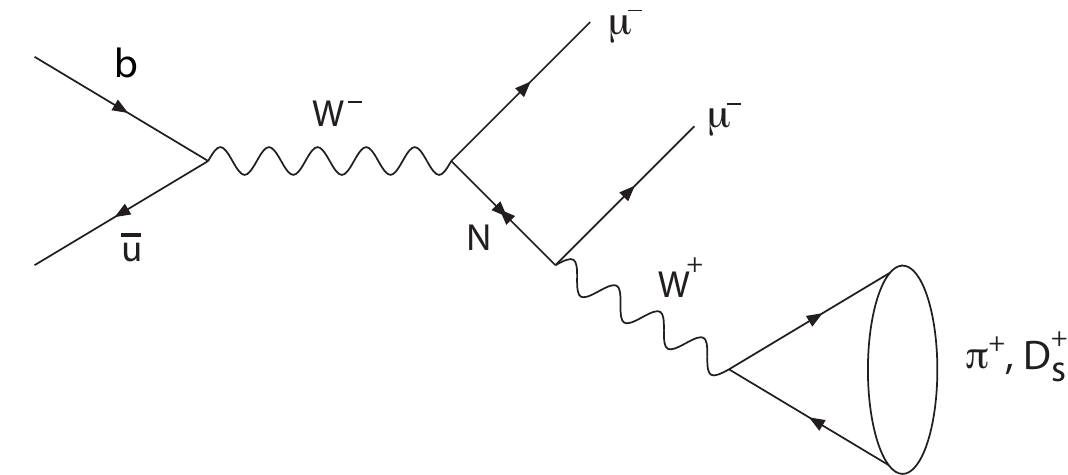}
      \end{center}
 \vspace{-.4mm}
\caption{(Feynman diagram for nuclear double-$\beta$ decay. (b) A related diagram for $B^-\to D^{+(*) }\mu^-\mu^-$ decays.}
\label{Majorana-annih}
\end{figure}

Majorana neutrinos with mass below that of the $B$ can be searched for directly. Consider the annihilation diagram shown in Fig.~\ref{Majorana-annih}. The process is similar to the one for $B^-\to\tau^-\overline{\nu}$ shown in Fig.~\ref{Btotaunu}, but here the neutrino being Majorana and on shell  can transform into a $\mu^-$ and a virtual $W^+$, which can materialize into hadrons. In particular the decay rate into $\pi^+$ and $D_s^+$ has been predicted as a function of the coupling between the heavy Majorana neutrino and the lepton, $|V_{\mu 4}|$ \cite{Majorana-thry-annih}. \footnote{Similar calculations have been performed for semi-leptonic decays  \cite{Majorana-thry-semi}, but these turn out to be less sensitive.}

\begin{figure}[htb] 
  \begin{center}
    \includegraphics*[width=0.7\textwidth]{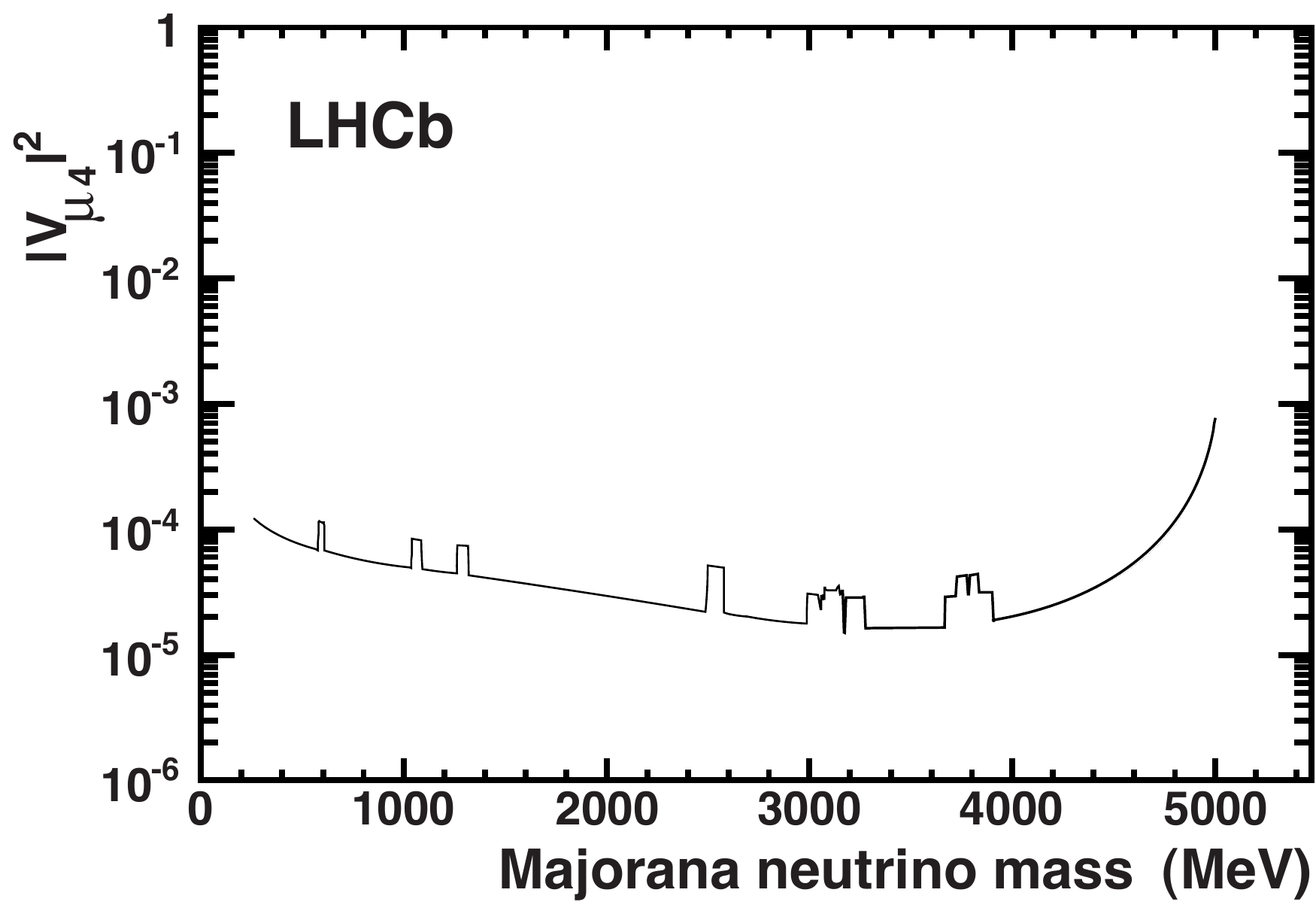}
      \end{center}
 \vspace{-.4mm}
\caption{Upper limits on $|V_{\mu 4}|$ at 95\% CL as a function of the Majorana neutrino mass derived from the $B^-\to \mu^-\mu^-\pi^+$ channel.}
\label{UL-Vmu4}
\end{figure}

LHCb has searched for these decays. They did not find any signals but have set the best experimental limits on $|V_{\mu 4}|$ as a function of the Majorana neutrino mass, for values just above the pion mass to just below the $B^-$ mass, shown in Fig.~\ref{UL-Vmu4} \cite{LHCb-Majorana}. Other searches have been carried out using like-sign dileptons at higher masses \cite{LHC-Majorana}.

\section{Conclusions}

Although there is no compelling evidence yet for NP,  Flavour Physics is very sensitive to potential effects at high mass scales. All NP theories must satisfy stringent experimental constraints.
LHCb has been very effective at dispelling effects with marginal statistical significance, although a few remain. Will some be established when precision increases?
Improving measurements in such decays $\overline{B}\to \overline{K}^*\mu^+\mu^-$, $\overline{B}_s^0\to\mu^+\mu^-$, $B^-\to \tau^-\overline{\nu}$, $\overline{B}\to D^{(*)}\tau^-\overline{\nu}$, $CP$ violation in $\overline{B}_s^0$ decays may show NP effects with increasing precision, and need to be aggressively pursued. Flavour based experiments also are looking for evidence that neutrinos are Majorana, and direct links to dark matter.
We are looking forward to defining the next theory beyond the SM either with our current and near term data or with new future facilities including Belle-II \cite{Super-Belle}, and the proposed LHCb upgrade \cite{LHCb-upgrade}.

\afterpage{\clearpage}
 \section*{Acknowledgements}
 I thank the U. S. National Science Foundation for support. My LHCb colleagues have provided a great intellectual environment conducive to thinking about this physics. I want to especially thank Marina Artuso and Liming Zhang for their help. My scientific secretary, Antonio Limosani, was very helpful in preparing my conference talk. I also thank the organizers, especially Geoffrey Taylor, Raymond Volkas, and Elisabetta Barberio who brought us to wonderful Melbourne.

\end{document}